\documentclass[aps,twocolumn,preprintnumbers,superscriptaddress,floatfix,pra,10pt]{revtex4-2}
\usepackage{bbm}
\usepackage{bm}
\usepackage{amsmath}
\usepackage{amssymb}
\usepackage{empheq}
\usepackage{graphicx}
\usepackage{mathrsfs}
\usepackage{amsfonts}
\usepackage{amsthm}
\usepackage{color}
\usepackage{multirow}
\usepackage{bigints}
\usepackage{txfonts}
\usepackage{float}
\usepackage{hyperref}
\hypersetup{
     unicode=false,              
     pdftoolbar=true,           
     pdfmenubar=true,         
     pdffitwindow=false,      
     pdfstartview={FitH},    
     pdftitle={My title},       
     pdfauthor={Author},     
     pdfsubject={Subject},   
     pdfcreator={Creator},   
     pdfproducer={Producer}, 
     pdfkeywords={keyword1} {key2} {key3}, 
     pdfnewwindow=true, 
     colorlinks=false,       
     linkcolor=red,           
     citecolor=green,        
     filecolor=magenta,    
     urlcolor=cyan           
}

\makeatletter \tolerance = 10000 \tolerance = 10000

\makeatother
\begin{document}

\title{Anderson localization transitions in disordered non-Hermitian systems with exceptional points}

\author{C. Wang}
\email[Corresponding author: ]{physcwang@tju.edu.cn}
\affiliation{Center for Joint Quantum Studies and Department of Physics, 
School of Science, Tianjin University, Tianjin 300350, China}
\author{X. R. Wang}
\email[Corresponding author: ]{phxwan@ust.hk}
\affiliation{Physics Department, The Hong Kong University of Science 
and Technology (HKUST), Clear Water Bay, Kowloon, Hong Kong}
\affiliation{HKUST Shenzhen Research Institute, Shenzhen 518057, China}

\date{\today}

\begin{abstract}
The critical exponents of continuous phase transitions of a Hermitian system depend on and only on its dimensionality and symmetries. This is the celebrated notion of the universality of continuous phase transitions. Here, we numerically study the Anderson localization transitions in non-Hermitian two-dimensional (2D) systems with exceptional points by using the finite-size scaling analysis of the participation ratios. At the exceptional points of either second-order or fourth-order, two non-Hermitian systems with different symmetries have the same critical exponent $\nu\simeq 2$ of correlation lengths, which differs from all known 2D disordered Hermitian and non-Hermitian systems. These feature is reminiscent of the superuniversality notion of Anderson localization transitions. In the symmetry-preserved and symmetry-broken phases, the non-Hermitian models with time-reversal symmetry and without spin-rotational symmetry, and without both time-reversal and spin-rotational symmetries, are in the same universality class of 2D Hermitian electron systems of Gaussian symplectic and unitary ensembles, where $\nu\simeq 2.7$ and $\nu\simeq 2.3$, respectively. The universality of the transition is further confirmed by showing that the critical exponent $\nu$ does not depend on the form of disorders and boundary conditions.
\end{abstract}

\maketitle

\section{Introduction}
\label{section1}

Disorder-induced quantum phase transitions from extended states to localized states, known as the Anderson localization transitions (ALTs)~\cite{pwanderson_pr_1958}, are a fundamental topic in wave physics. ALTs can be divided into different classes. Each class has a set of specific critical exponents that depend only on the dimensionality and symmetries of the class and not on the details of the disordered Hamiltonians. This is the notion of the universality of continuous phase transitions~\cite{palee_rmp_1985,bkramer_rpp_1993,fevers_rmp_2008}. In Hermitian cases, disordered metals are classified into Gaussian orthogonal, unitary, and symplectic ensembles according to time-reversal and spin-rotational symmetries. Disordered Hermitian metals are classified into Gaussian unitary ensemble if they do not have time-reversal symmetry (TRS), Gaussian orthogonal ensemble if they have both TRS and spin-rotational symmetry, and Gaussian symplectic ensemble if they have the TRS, but without the spin-rotational symmetry. Different symmetry classes have different critical exponents near the ALTs that depend only on their dimensionality. For example, disordered two-dimensional (2D) electron gases exhibit the integer quantum Hall effect when TRS is broken~\cite{hlevine_npb_1984,bhuckestein_rmp_1995}. In the presence of weak spin-orbit interactions, the critical exponent of correlation length is $\nu\simeq 2.3$~\cite{cwang_prl_2015,ysu_scirep_2016,cwang_prb_2017}, while the same gases without a magnetic field, such that the systems belong to the Gaussian symplectic class, have $\nu\simeq 2.7$~\cite{snevangelou_prl_1995,rmerkt_prb_1998,yasada_prl_2002}.
\par

Hamiltonians of all open systems are ubiquitously non-Hermitian, and non-Hermiticity leads to fundamentally different phenomena in non-Hermitian systems from their counterparts of Hermitian systems in all aspects, including the topological properties~\cite{telee_prl_2016,fkkunst_prl_2018,syao_prl_2018,cwang_prb_2022} and the Anderson localizations~\cite{nhatano_prl_1996,cwang_prb_2020,yhuang_prb_2020}. For example, while the critical dimension of ALTs for Hermitian systems is two~\cite{bkramer_rpp_1993}, extended states can appear in one-dimensional systems with non-Hermiticities~\cite{nhatano_prl_1996}. Likewise, disordered non-Hermitian systems can be classified by their symmetries, which leads to a 38-fold symmetry classification~\cite{dbernard_arxiv_2020,kkawabata_prx_2019}. Based on the 38-fold classification, previous works numerically investigate the universality of ALTs of some symmetry classes~\cite{xluo_prl_2021,kkawabata_prl_2021,xluo_prb_2021,xluo_prresearch_2022}.   
\par   

Noticeably, non-Hermitian systems with specific symmetries [e.g., parity-time symmetry (PTS)~\cite{cmbender_prl_1998} or pseudo-Hermitian symmetry~\cite{amostafazadeh_jmp_2002}] display exceptional points (EPs), where the right eigenstates coalesce and become orthogonal to the corresponding left ones~\cite{wdheiss_jpa_2012}. EPs have recently attracted enormous attention because of their exotic properties and potential applications in spintronics~\cite{hyang_prl_2018}, electronics~\cite{zxiao_prl_2019}, photonics~\cite{skozdemir_natmater_2019}, and optics~\cite{mamiri_science_2019}. However, many theoretical efforts have focused on the topological properties of EPs in the clean limit~\cite{smalzard_prl_2015,dleykam_prl_2017,yxu_prl_2017,kkawabata_prl_2019,xxzhang_prl_2020}, and a systematic study of ALTs of disordered non-Hermitian systems with EPs is lacking.
\par

Here, we investigate the ALTs of two 2D non-Hermitian systems with different symmetries and with EPs at fixed points in the complex-energy plane. Based on the finite-size scaling analysis of participation ratios, we find that critical exponents of ALTs at either the second-order or fourth-order EPs of different symmetry classes (class AIII or class  DIII$+{\cal S}_{-+}$) are identical within numerical errors ($\nu=2$) and are distinctive in those of any known symmetry classes. These findings indicate that ALTs at EPs in different symmetry classes form one universality class, a behavior which, together with a different critical exponent $\nu=2$, is evocative of a new superuniversality class. Besides, ALTs in the symmetry-preserved and symmetry-broken phases, where the energy spectra are real and complex, respectively, belong to the same universality class. Now, the critical exponents are $\nu\simeq 2.7$ and $2.3$, depending on whether TRS is presented, and agree with those of the 2D symplectic class~\cite{snevangelou_prl_1995,rmerkt_prb_1998,yasada_prl_2002} and unitary class with spin-orbit interactions in Hermitian systems~\cite{cwang_prl_2015,ysu_scirep_2016,cwang_prb_2017}, respectively. To further substantiate the universality, we numerically show that the types of disorders and boundary conditions do not affect the critical exponents at EPs.   
\par

The paper is organized as follows: The lattice models and the methods are described in Secs.~\ref{section2} and \ref{section3}, respectively. Numerical results are presented in Sec.~\ref{section4}, followed by a discussion in Sec.~\ref{section5} and a conclusion in Sec.~\ref{section6}.

\section{Models and symmetries}
\label{section2}

We study non-Hermitian Hamiltonians $H$ that transform under certain transformation operators $\mathcal{O}$ as $[\mathcal{O},H]_{\zeta=\pm 1}=\mathcal{O}H-\zeta H\mathcal{O}=0$. If $\mathcal{O}$ is the product of parity and time-reversal operators or a pseudo-Hermitian symmetry operator, eigenenergies of $H$ are real or form pairs of complex-conjugate numbers, i.e., $\epsilon\in \mathbb{R}$ or $( \epsilon,\epsilon^\ast )$~\cite{cwang_prbl_2022}. For a Bloch Hamiltonian $h(\bm{k})$, the two symmetries can be written as $[u_{\mathcal{PT}}\mathcal{K},h(\bm{k})]_{\zeta=1}=0$ and $[u_q\eta,h(\bm{k})]_{\zeta=1}=0$ with $u_{\mathcal{PT}}$ and $u_q$ being unitary operators and $\mathcal{K}$ and $\eta$ being complex-conjugate and Hermitian-conjugate operators, respectively. The critical point separating the real-energy and complex-energy spectra is thus the EP~\cite{hyang_prl_2018}. One should not be confused PTS with TRS defined by $[u_{\mathcal{T}}\mathcal{K}\mathcal{K}_k,h(\bm{k})]_{\zeta=1}=0$ with $\mathcal{K}_k$ changing $\bm{k}$ to $-\bm{k}$ in the momentum space. Interpretations of PTS and TRS are given in Appendix~\ref{sec_1_1}.
\par

To investigate ALTs at an EP, an accurate trace of its location is required. Such process is easy for clean systems where closed-form solutions of eigenvalues are available but is difficult for disordered systems where the positions of EPs may be random. Here, our strategy is to consider non-Hermitian systems with an additional parity-particle-hole symmetry (PPHS), defined by $[u_{\mathcal{P}P} \lambda,h(\bm{k})]_{\zeta=-1}=0$ with $u_{\mathcal{P}P}$ being a unitary operator and $\lambda$ being the transpose operator, such that eigenvalues are in pairs of $(-\epsilon,\epsilon )$. This constraint, together with PTS, leads to a cross shape of eigenvalue distributions on the complex-energy plane whose EPs are fixed at the origin, see Appendix~\ref{sec_1_2}. It should be mentioned that PPHS is different from particle-hole symmetry (PHS) which is defined by $[u_{\mathcal{P}}\lambda\mathcal{K}_k,h(\bm{k})]_{\zeta=-1}=0$ of $u_{\mathcal{P}}$ being a unitary matrix. A non-Hermitian system with PPHS means it is invariant under a transformation of the product of the parity inversion and the particle-hole operations. The definition of PPHS is given in Appendix~\ref{sec_1_1}.
\par

Under these considerations, we study two non-Hermitian tight-binding models on square lattices of size $L\times L$ with lattice constant $a=1$ of different symmetries. The first one reads
\begin{equation}
\begin{gathered}
H_1=\sum_{\bm{i}}c^\dagger_{\bm{i}}[(w_{\bm{i}}+iu_{\bm{i}})\sigma_1+i\kappa\sigma_3]c_{\bm{i}}\\
-\sum_{\bm{i}}\left[ \dfrac{i\alpha}{2}c^\dagger_{\bm{i}}(\sigma_2 c_{\bm{i}+\hat{x}}-\sigma_1 c_{\bm{i}+\hat{y}})+H.c. \right].
\end{gathered}\label{eq1}
\end{equation}
Here, $c^\dagger_{\bm{i}}$ ($c_{\bm{i}}$) is the creation (annihilation) operator of a spinor on lattice site $\bm{i}$. $\sigma_{0,1,2,3}$ are the identity and Pauli matrices acting on the spin space. $\hat{x}$ and $\hat{y}$ are the unit vectors along the $x$ and $y$ directions, respectively. $\kappa$ and $\alpha$ are real positive constants. Disorders are modelled by the on-site potential $(w_{\bm{i}}+iu_{\bm{i}})\sigma_1$, where $w_{\bm{i}}$ and $u_{\bm{i}}$ are real numbers and distribute uncorrelatively and uniformly in the range of $[-W/2,W/2]$. 
\par

For $w_{\bm{i}}=u_{\bm{i}}=0$, $H_1$ can be block-diagonalized with the Bloch Hamiltonian $h_1(\bm{k})=\alpha\sin k_2\sigma_1-\alpha\sin k_1\sigma_2+i\kappa\sigma_3$. $h_1(\bm{k})$ has PTS with $[\sigma_1\mathcal{K},h_1(\bm{k})]_{\zeta=1}=0$. The appearance of EPs can be seen from the eigenvalues of $h_1(\bm{k})$: $\epsilon^{\pm}_{1}=\pm(\alpha^2(\sin^2 k_1+\sin^2 k_2)-\kappa^2)^{1/2}$, which are real for $\alpha^2(\sin^2 k_1+\sin^2 k_2)>\kappa^2$ or come as complex pairs for $\alpha^2(\sin^2 k_1+\sin^2 k_2)<\kappa^2$. The domain with real-energy is termed as the symmetry-preserved phase; otherwise known as the symmetry-broken phase. Therefore, for $\alpha>\kappa/\sqrt{2}$, the two phases are separated by an EP locating at $\epsilon^{\pm}_{1}=0$ where $\alpha^2(\sin^2 k_1+\sin^2 k_2)=\kappa^2$.  
\par

\begin{figure}[htbp]
\centering
\includegraphics[width=0.45\textwidth]{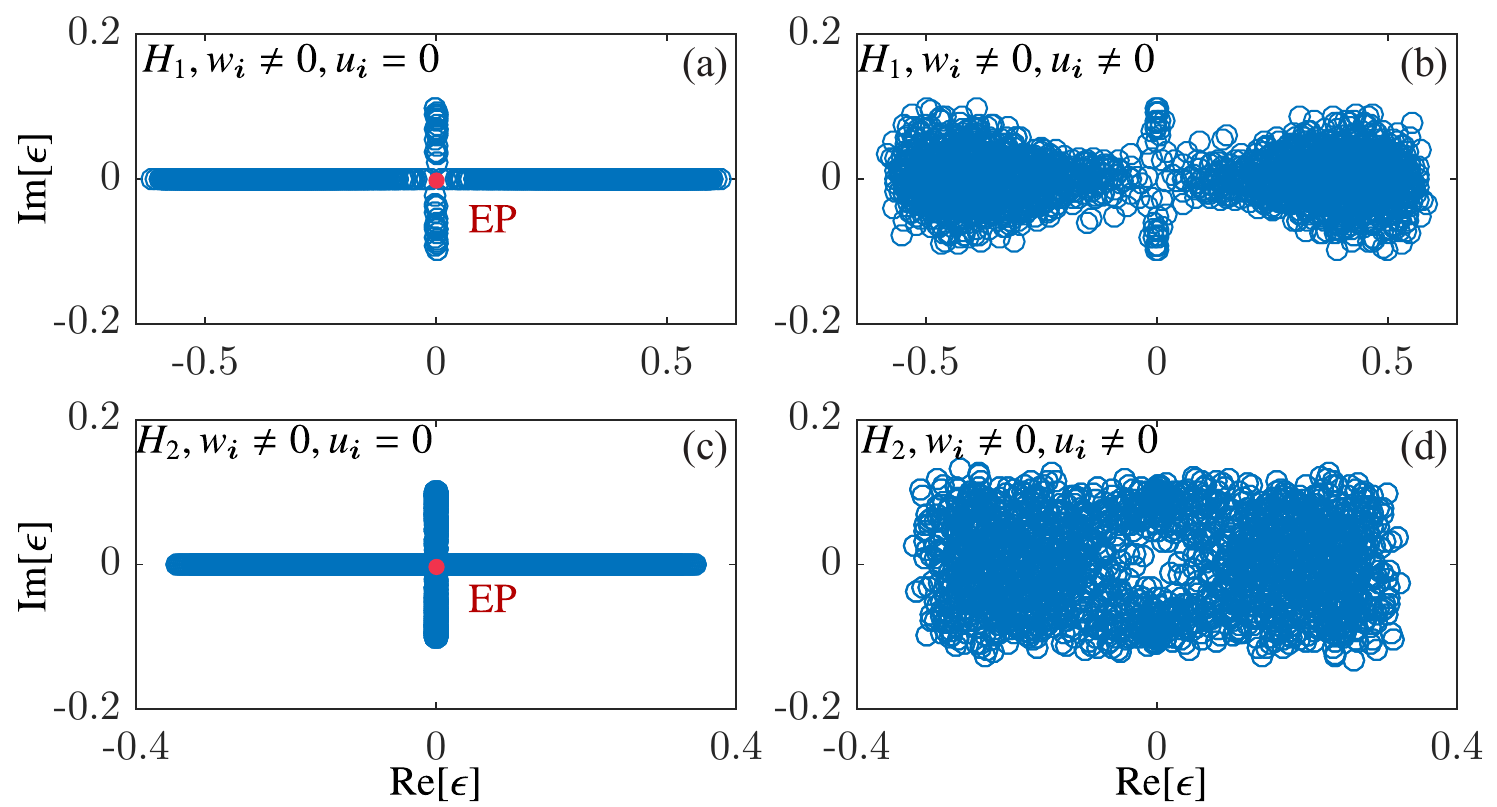}
\caption{(a) Eigenenergy distribution in the complex-energy plane of $H_1$ for $\alpha=0.2$, $\kappa=0.1$, $L=30$, $u_{\bm{i}}=0$ and $w_{\bm{i}}\in [-W/2,W/2]$ with $W=0.3$. (b) Same as (a) but for $u_{\bm{i}}=w_{\bm{i}}$. (c) Same as (a) but for $H_2$. (d) Same as (a) but for $H_2$ and $u_{\bm{i}}=w_{\bm{i}}$. $10^2$ samples are used for each plot. Red dots denote the EP positions in (a,c). No EP is observed for (b,d) since the disorder terms $i u_{\bm{i}}\sigma_1$ and $i u_{\bm{i}}\tau_2\sigma_0$ break PTS.}
\label{fig1}
\end{figure}

In addition to PTS, $h_1(\bm{k})$ has PPHS as well, i.e., $[i\sigma_2\lambda,h_1(\bm{k})]_{\zeta=-1}=0$, such that $\epsilon^{\pm}_{1}$ are symmetric to the origin of the complex-energy plane. Disorders breaks lattice-translational symmetry but preserves PPHS since $[ i\sigma_2\lambda,(w_{\bm{i}}+iu_{\bm{i}})\sigma_1 ]_{\zeta=-1}=0$. Differently, PTS is preserved only if $u_{\bm{i}}=0$ since $[\sigma_1\mathcal{K}, w_{\bm{i}}\sigma_1]_{\zeta=1}=0$ and $[\sigma_1\mathcal{K}, i u_{\bm{i}}\sigma_1]_{\zeta=1}\neq 0$. Hence, for $w_{\bm{i}}\neq 0$ and $u_{\bm{i}}=0$, $H_1$ has both PTS and PPHS whose eigenvalues are in the cross region of the complex-energy plane and the EP is at the origin; while, for $w_{\bm{i}}\neq 0$ and $u_{\bm{i}}\neq 0$, PTS is broken, and no EP is found, as shown in Figs.~\ref{fig1}(a) and (b).
\par 

It is worth classifying the model $H_1$ within the framework of Altland-Zirnbauer (AZ) classification. In this symmetry classification, one require to consider TRS, PHS, particle-hole symmetry$^\dagger$ (PHS$^\dagger$), time-reversal symmetry$^\dagger$ (TRS$^\dagger$), chiral symmetry (CS), and sub-lattice symmetry (SLS), rather than PTS and PPHS, see Appendix~\ref{sec_1_3} for more details. For $H_1$ with $w_{\bm{i}}\neq 0,u_{\bm{i}}=0$, TRS, PHS, PHS$^\dagger$, TRS$^\dagger$, and SLS are broken but CS is preserved. Therefore, $H_1$ of $w_{\bm{i}}\neq 0,u_{\bm{i}}=0$ belongs to class AIII. On the other hand, $H_1$ with $w_{\bm{i}}\neq 0,u_{\bm{i}}\neq 0$ breaks all symmetries of the AZ classification and belongs class A. We summarize these results in Table~\ref{tab_2} of Appendix~\ref{sec_1_3}.
\par

The second model reads
\begin{equation}
\begin{gathered}
H_2=\sum_{\bm{i}}c^\dagger_{\bm{i}}[(w_{\bm{i}}+i u_{\bm{i}})\tau_2\sigma_0+i\kappa\tau_3\sigma_3]c_{\bm{i}}\\
-\sum_{\bm{i}}\left[ \dfrac{i\alpha}{2}c^\dagger_{\bm{i}}(\tau_0\sigma_2 c_{\bm{i}+\hat{x}}-\tau_0\sigma_1 c_{\bm{i}+\hat{y}})+H.c. \right],
\end{gathered}\label{eq2}
\end{equation}
where $\tau_{0,1,2,3}$ are the identity and the Pauli matrices acting on the pseudo-spin space. Disorders are modelled by the on-site term $(w_{\bm{i}}+i u_{\bm{i}})\tau_2\sigma_0$. The Bloch Hamiltonian of $H_2$ is $h_2(\bm{k})=\alpha\sin k_1\tau_0\sigma_2-\alpha\sin k_2\tau_0\sigma_1+i\kappa\tau_3\sigma_3$, which has both PTS and PPHS, i.e., $[\tau_3\sigma_1\mathcal{K},h_2(\bm{k})]_{\zeta=1}=[i\tau_0\sigma_2\lambda,h_2(\bm{k})]_{\zeta=-1}=0$. Therefore, the complex-energy spectra of $h_2(\bm{k})$ are symmetric to the origin of the complex-energy plane: $\epsilon^{\pm,s}_{2}=\pm(\alpha^2(\sin^2 k_1+\sin ^2 k_2)-\kappa^2)^{1/2}$ with the EP at $\epsilon=0$ and $s=1,2$ standing for a two-fold degeneracy. With disorders, $w_{\bm{i}}\tau_2\sigma_0$ ($i u_{\bm{i}}\tau_2\sigma_0$) preserves (breaks) PTS. Hence, $H_2$ with $w_{\bm{i}}\neq 0$ and $u_{\bm{i}}=0$ has an EP at $\epsilon=0$, and no EP is expected for $w_{\bm{i}}\neq 0$ and $u_{\bm{i}}\neq 0$, see Figs.~\ref{fig1}(c) and (d).  Recall that the energy spectrum of $H_2$ is two-fold degenerated. Consequently, the EP shown in Fig.~\ref{fig1}(c) is forth-order, different from the second-order EP in Fig.~\ref{fig1}(a).
\par

In addition to PTS and PPHS, $H_2$ has TRS, PHS, TRS$^\dagger$, and PHS$^\dagger$ when $w_{\bm{i}}\neq 0,u_{\bm{i}}=0$ and belongs to class DIII$+{\cal S}_{-+}$. Therefore, for $u_{\bm{i}}=0$, $H_2$ (class DIII$+{\cal S}_{-+}$) and $H_1$ (class AIII) belong to different symmetry classes in the AZ classification even though both of them have EPs. Differently, TRS and PHS$^\dagger$ are broken if $u_{\bm{i}}\neq 0$, and $H_2$ belongs to class DIII$+{\cal S}_{-}$. A detailed analysis of symmetries is given in Appendix~\ref{sec_1_3} and the results are summarized in Table~\ref{tab_2}. 
\par

\section{Numerical methods}
\label{section3}

Complex mobility edge (complex energy at an ALT) can be numerically identified from the finite-size scaling analysis of the  participation ratio $p_2$ of a state with energy $\epsilon$ defined as $p_2(\epsilon)=\langle (\sum_{\bm{i}}|\psi_{\bm{i},\epsilon}|^{4})^{-1}\rangle$. Here, $\psi_{\bm{i},\epsilon}$ is the normalized wave function amplitude of a right eigenstate, and $\langle\cdots\rangle$ denotes the ensemble-average. $p_2$ scales with the system length as $p_2\propto L^2$ for extended states and approaches a constant for localized states~\cite{cwang_prl_2015}. If there is an ALT for a given state at a critical disorder $W_c$, $p_2$ near $W_c$ behaves as
\begin{equation}
\begin{gathered}
p_2=L^D[f(L/\xi)+\phi L^y \tilde{f}(L/\xi)]
\end{gathered}\label{eq3}
\end{equation}
with $D\in [0,2]$ being the fractal dimension \cite{wang_pra_1989} of the critical wave function, $\phi$ being a positive constant, and $y<0$ being the exponent of irrelevant scaling parameters. $\xi$ is the correlation length and diverges as $\xi\propto |W-W_c|^{-\nu}$ near $W_c$ with $\nu>0$ being the universal critical exponent. The validity of the single-parameter scaling Eq.~\eqref{eq3} has been confirmed in both Hermitian~\cite{jhpixley_prl_2015,cwang_prb_2019} and non-Hermitian systems~\cite{cwang_prb_2020,xluo_prresearch_2022}. ALTs in the same universality class have identical critical exponents. 
\par

In our approach, $p_2(\epsilon)$ is numerically computed through the exact diagonalizations by using the KWTANT package~\cite{kwant} and the SciPy library~\cite{scipy} on Python. Then, a chi-square fit of $p_2$ to the scaling function Eq.~\eqref{eq3} is performed by a polynomial expansion~\cite{wchen_prb_2019} from which we obtain $W_c$, $\nu$, $D$, $\phi$, $y$, and the unknown scaling functions $f(x)$ and $\tilde{f}(x)$. All fittings have satisfactory goodness-of-fits $Q>0.01$. Curves $Y_L(W)=p_2 L^{-D}- \phi L^y \tilde{f}(L/\xi)$ for different sample size $L$ are used to identify an ALT by the following criteria: (i) $Y_L(W)$ increases (decreases) with $L$ for extended (localized) states. (ii) $Y_L(W)$ for different $L$ cross each other at $W_c$. (iii) $Y_L(W)$ for different $L$ collapse to a smooth scaling function $f(L/\xi)$ near $W_c$.
\par

Note that $H_1$ and $H_2$ cannot be diagonalized at the EP~\cite{wdheiss_jpa_2012}. Instead, we calculate $p_2(\tilde{\epsilon})$ with $\tilde{\epsilon}$ being the nearest eigenvalue to the EPs and assume $p_2(0)=p_2(\tilde{\epsilon})$. This approximation should be valid in the thermodynamic limit $L\to\infty$ and for $\tilde{\epsilon}$ extremely close to EPs, see Appendix~\ref{sec_2}. For finite-size systems, there should be a critical length $\tilde{L}$ above which the critical exponent $\nu$, obtained by scaling analysis of $p_2(\tilde{\epsilon})$, keeps unchanged within numerical errors. Hence, the approximation should be acceptable for $L>\tilde{L}$. We find $\tilde{L}=80$ for $H_1$ and $H_2$. Numerical evidence is given in Appendix~\ref{sec_2}. 
\par

\section{Results}
\label{section4}  

\begin{figure}[htbp]
\centering
\includegraphics[width=0.45\textwidth]{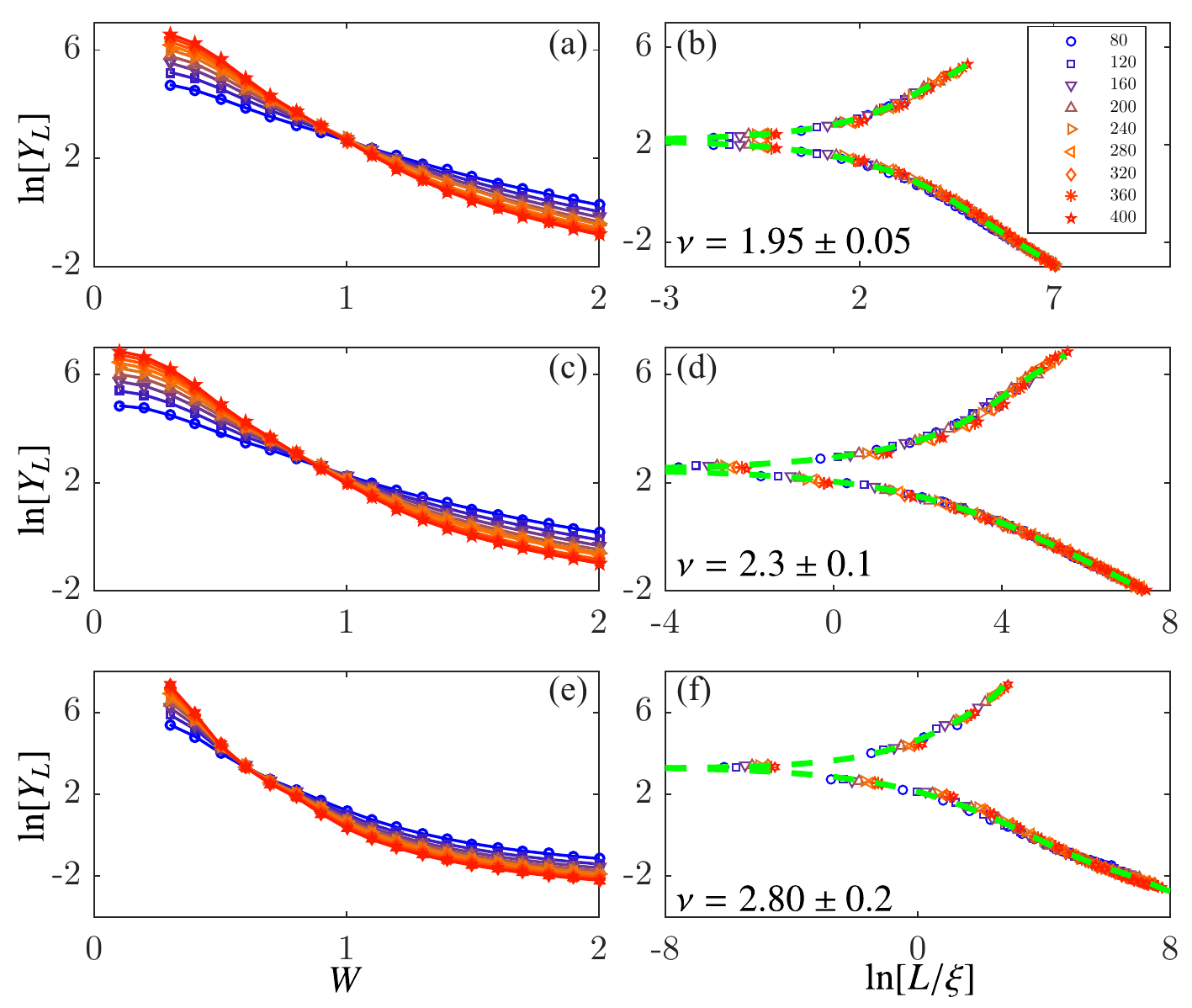}
\caption{(a) $\ln[Y_L(W)]$ of $H_1$ at $\epsilon=0$ (the EP) and for $L=80,120,\cdots,400$. Here, $\alpha=0.2$, $\kappa=0.1$, and $u_{\bm{i}}=0$. (b) $\ln [Y_L]$ v.s. $\ln [L/\xi]$ for data of (a). Each data is average over more than $10^3$ samples. (c,d) Same as (a,b) but at $\epsilon=-0.2$ (symmetry-preserved phase). (e,f) Same as (a,b) but for $\epsilon=-0.2+i0.01$ and $w_{\bm{i}}=u_{\bm{i}}$. Other fitting parameters are given in Appendix~\ref{sec_3}.}
\label{fig2}
\end{figure}

Let us first consider $H_1$ with $\alpha=0.2$, $\kappa=0.1$, and $u_{\bm{i}}=0$. The system has an EP at $\epsilon=0$ as shown in Fig.~\ref{fig1}(a). Figure~\ref{fig2}(a) shows $\ln[Y_L(W)]$ of $\epsilon=0$ for various $L$ ranging from 80 to 400. Here, curves of different $L$ cross at a single point $W_c=0.85\pm 0.05$, and states of $W<W_c$ ($W>W_c$) are extended (localized) because $Y_L$ increases (decreases) with $L$. Finite-size scaling analysis yields $\nu=1.95\pm 0.07$, which is different from any known critical exponents of disordered non-Hermitian systems, indicating that the ALT belongs to an unknown universality class. Data near the critical point collapse on a single smooth scaling function with two branches for the extended and localized states, see Fig.~\ref{fig2}(b). 
\par

For the state at energy $\epsilon=-0.2$ in the symmetry-preserved phase, an ALT occurs at $W_c=0.93\pm 0.05$, see Fig.~\ref{fig2}(c). The beautiful scaling function shown in Fig.~\ref{fig2}(d) substantiates the criticality of the transition. The fitting suggests that $\nu=2.3\pm 0.1$ which equals to that of Hermitian spinful Gaussian unitary ensemble~\cite{cwang_prb_2017}, where TRS and spin-rotational symmetries are broken. This is because the non-Hermitian systems in the symmetry-preserved phase behave as the Hermitian systems and the disorder term $w_{\bm{i}}\sigma_1$ breaks TRS, see Appendix~\ref{sec_1_3}. Hence, the ALT shown in Figs.~\ref{fig2}(c) and (d) belongs to the same universality class of the spinful Gaussian unitary class~\cite{cwang_prb_2017}.
\par

To further confirm the universality shown in Figs.~\ref{fig2}(a)-(d), we carry out numerical calculations of the dimensionless conductance $g_L$ based on the transfer matrix method and perform the corresponding finite-size scaling analyse for $\epsilon=-0.01$ (near the EP) and $\epsilon=-0.2$ (the symmetry-preserved phase) of $H_1$ with $u_{\bm{i}}=0$, see Appendix~\ref{sec_4_1}. The obtained critical exponents are $\nu=2.05\pm 0.07$ and $2.33\pm 0.05$, respectively, which accord with those based on participation ratios within numerical errors. Besides, the critical exponent for the symmetry-preserved phase is also close to that for 2D Hermitian spinful Gaussian unitary ensemble~\cite{cwang_prb_2017}, which is consistent with a recent work that estimates an equivalent mapping between the universality of class AIII for Hermitian systems and class A for non-Hermitian systems~\cite{xluo_prresearch_2022}.
\par

An ALT also occurs at $W_c=0.83\pm 0.02$ for the state at $\epsilon=0.08i$ in the symmetry-broken phase, as shown in Appendix~\ref{sec_4_3}. The calculated critical exponent, $\nu=2.32\pm 0.02$, equals to that of $\epsilon=-0.2$ within numerical errors, indicating that they have the same universality. The same universality of the symmetry-preserved and symmetry-broken phases can be understood as follows: The symmetry-preserved phase of $H_1$ with $u_{\bm{i}}=0$ can be mapped into the symmetry-broken phase under the transformation $H_1\to iH_1$ due to the presence of both PTS and PPHS. The critical exponents of $H_1$ and $iH_1$ should be the same since this mapping exchange only real and imaginary axes without changing their eigenfunctions. 
\par

For $u_{\bm{i}}\neq 0$, PTS is broken, and the EP disappears. Below, we set that both $u_{\bm{i}}$ and $w_{\bm{i}}$ uniformly distribute in $[-W/2,W/2]$. For a state at $\epsilon=-0.2+i0.01$, an ALT can be identified at $W_c=0.62\pm 0.03$, see Figs.~\ref{fig2}(e) and (f). The critical exponent is $\nu=2.79\pm 0.02$ that is significantly different from those in Figs.~\ref{fig2}(a)-(d), i.e., a distinguished universality class. From the best of our knowledge, there is no estimation of $\nu$ for non-Hermitian systems with PPHS $[\sigma_2\lambda,H_1 ]_{\zeta=1}=0$ in 2D. Remarkably, the obtained $\nu$ is very close to those of classes AIII, CII$^\dagger$, and DIII within the AZ classification~\cite{xluo_prresearch_2022}.  
\par

\begin{figure}[htbp]
\centering
\includegraphics[width=0.45\textwidth]{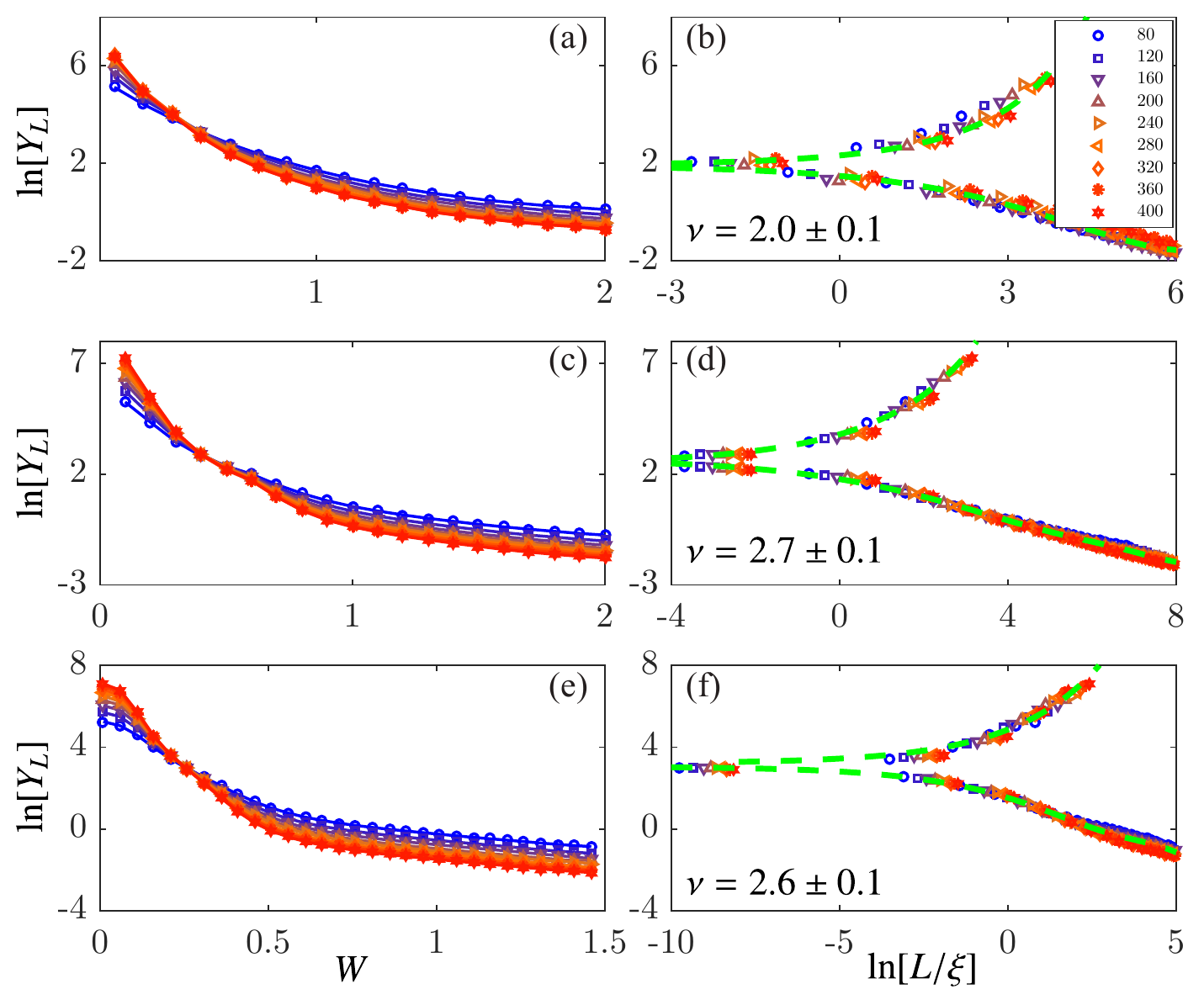}
\caption{(a) $\ln[Y_L]$ as a function $W$ of $H_2$ with $\epsilon=0$, $\alpha=0.2$, $\kappa=0.1$, and $L=80,120,\cdots,280$. (b) $\ln [Y_L(\ln[L/\xi])]$ for data of (a). (c,d) Same as (a,b) but at $\epsilon=-0.2$ (symmetry-preserved phase). (e,f) Same as (a,b) but at $\epsilon=-0.2+0.01i$ for $u_{\bm{i}}$ distributing uniformly in the range of $[-0.1,0.1]$. Other fitting parameters are given in Appendix~\ref{sec_3}.}
\label{fig3}
\end{figure}

Now, let us turn to $H_2$ with $\alpha=0.2,\kappa=0.1,u_{\bm{i}}=0$ that belongs to class DIII$+{\cal S}_{-+}$. Similar to $H_1$, there is an ALT at $W_c=0.65\pm 0.03$ for states near the EP, see Figs.~\ref{fig3}(a) and (b). The critical exponent $\nu=2.0\pm 0.1$, the same as that of $H_1$ at the EP within numerical errors even though $H_1$ and $H_2$ with $u_{\bm{i}}=0$ belong to different symmetry classes (class A and DIII$+{\cal S}_{-+}$, respectively) and the orders of EPs are different. Our results, presented in Figs.~\ref{fig2} and \ref{fig3}, suggest that ALTs for the states at EPs have the same critical exponent $\nu\simeq 2$. This notion of ``superuniversality'' reminisces similar concept in disordered Hermitian superconducting systems~\cite{iagruzberg_prb_2005}.
\par

To understand critical properties of ALTs for the symmetry-preserved states of $H_2$, state of $\epsilon=-0.2$ is studied. As shown in Figs.~\ref{fig3}(c) and (d), an ALT occurs at $W_c=0.45\pm 0.05$ with $\nu=2.79 \pm 0.05$ which is significantly larger than that shown in Figs.~\ref{fig2}(c) and (d), but the same as that of 2D Gaussian sympletic ensembles in Hermitian random matrices~\cite{yasada_prl_2002}. Different from $H_1$ that breaks TRS, $H_2$ is invariant under the time-reversal transformation of $\Theta=\tau_3\sigma_2\mathcal{K}$, i.e., $[\Theta,H_2]_{\zeta=1}=0$. This explains why the critical exponents of the symmetry-preserved states of $H_2$ fall into the universality class of $\nu\simeq 2.7$~\cite{yasada_prl_2002} for the Hermitian time-reversal-invariant systems with $\Theta^2=-I$.
\par

Interestingly, $H_2$ of $u_{\bm{i}}\neq0$ and $w_{\bm{i}}\neq0$, whose eigenvalues are complex, has different critical exponent as those above and thus belongs to a different universality class. This claim is derived from data of $\ln [Y_L(\epsilon=-0.2+i0.01)]$ for $u_{\bm{i}}$ distributing uniformly in the range of $[-0.1,0.1]$, as shown in Figs.~\ref{fig3}(e) and (f), where an ALT happens at $W_c=0.26\pm 0.01$ with $\nu=2.56\pm 0.03$. Thus, for systems without EPs, its universality class depends on symmetries, the same as their Hermitian counterparts. 
\par

\section{Discussions}
\label{section5}

\emph{Fractal dimension.$-$}Equation~\eqref{eq3} says $p_2\propto L^{D}$ at criticality, in contrast to $p_2\propto L^2$ for extended states and $p_2\propto L^0$ for localized states. The fractal dimension $D$ is universal according to the renormalization-group theory of the $\sigma$ model in $2+\epsilon$ dimensions~\cite{fwegner_zpb_1980}. At EPs, we obtain $D\simeq 1$ within numerical errors (see Tables in Appendix~\ref{sec_3}) that supports this argument. 
\par

$D$ relates to the spectral compressibility $\chi$ characterizing fluctuations of the energy level number $n$ in an energy window near the criticality, i.e., $\text{var}(n)=\chi\langle n\rangle$. For Hermitian systems, the following relation between $\chi$ and $D$ holds: $\chi=(d-D)/(2d)$~\cite{fevers_prl_2000}. Therefore, our results suggest a universal $\chi=1/4$ for EPs. In an early work, we have shown that the nearest-neighbor level-spacing distributions follow some universal functions near EPs~\cite{cwang_prbl_2022}. However, how to generate the concept of the spectral compressibility and test the correctness of $\chi=1/4$ remains unclear and deserves further studies.
\par

\emph{More evidence for the superuniversality at EPs.$-$}The meaning of universality requires the independence of $\nu$ on the boundary conditions and the forms of disorders~\cite{fevers_rmp_2008}. Since open boundary conditions and uniform distributions of random numbers are used in Figs.~\ref{fig2} and~\ref{fig3}, studies with periodical boundary condition and disorders of the normal-distribution are carried out to test the universality of $\nu$ at EPs, see Appendices~\ref{sec_4_3} and \ref{sec_4_4}. Indeed, the same critical exponents for EPs, $\nu\simeq 2$, is obtained in all cases.
\par

\emph{Skin effect.$-$}The non-Hermiticity itself can also lead to localizations of waves. A phenomenon is known as the non-Hermitian skin effect where the wave functions localize at the boundary of systems for some specific non-Hermiticities~\cite{syao_prl_2018}. However, our models, Eqs.~\eqref{eq1} and \eqref{eq2} with PTS, do not suffer from the skin effect such that all localizations shown here are due to disorders rather than non-Hermiticities, see Appendix~\ref{sec_5}.
\par

\emph{Materials relevance.$-$}The participation ratio $p_2(\epsilon)$ of a non-Hermitian Hamiltonian $H$ with a right eigenenergy $\epsilon$ is the same as $p_2(\epsilon-i\gamma I)$ of $H-i\gamma I$ with $I$ being the identity matrix of the same dimension of $H$. For $H\phi_\epsilon=\epsilon\phi_\epsilon$, one can easy to find that $(H-i\gamma I)\phi_\epsilon=(\epsilon-i\gamma)\phi_\epsilon$. By definition, $p_2(\epsilon)=p_2(\epsilon-i\gamma)$. Hence, additional on-site potentials $-i\gamma \sigma_0$ and $-i\gamma\sigma_0\tau_0$ to Hamiltonians Eq.~\eqref{eq1} and \eqref{eq2} do not change the participation ratios, as well as the universality.
\par

Note that the effective $\bm{k}\cdot\bm{p}$ Hamiltonian of $h_1$ near $\bm{k}=(0,0)$ reads $h_1(\bm{p})=\alpha(\bm{p}\times\bm{\sigma})\cdot\hat{z}+i\kappa\sigma_3$, which, together with a non-local loss $-i\gamma\sigma_0$ with $\gamma>\kappa$, which does not affect the criticality, describes a Rashba spin-orbit coupling with different spin lifetimes. Possible physical realizations of $H_1$ are ferromagnetic semiconductors such as MnGaAs and other III-V host materials~\cite{nnagaosa_rmp_201} with a spin-dependence impurity $w_{\bm{i}}\sigma_1$ that breaks TRS. Likewise, $H_2$ can be treated as the 2D electron gases with Rashba spin-orbit coupling and different lifetimes of spins/pseudo-spins, but the disorders $w_{\bm{i}}\tau_2\sigma_0$ are spin-independent.
\par

In addition to electronic systems, EPs present in many other systems, including lasers~\cite{gharari_science_2018,mabandres_science_2018}, micro-cavities~\cite{bpeng_natphys_2014}, electrics~\cite{vvkonotop_rmp_2016}, and magnonics~\cite{hyang_prl_2018}, to name a few. Disorders can be artificially induced in such systems such that the ALTs of the EPs in these systems can be experimentally studied in principle. In Appendix~\ref{sec_6}, a laser cavity network is proposed as a possible experimental verification of the numerical results presented in this paper.
\par

\section{Conclusion}
\label{section6}

In summary, ALTs at EPs of two non-Hermitian systems of different symmetries have the same critical exponent $\nu\simeq 2$, independent of the forms of disorders and the boundary conditions. This strongly suggests that ALTs at EPs of non-Hermitian systems belong to a new universality class that may depends only on dimensionality. Besides, the universality of ALTs of the symmetry-preserved and symmetry-broken phase is the same as their Hermitian counterpart and depends on the presence of their symmetries. 
\par

\begin{acknowledgments}

This work is supported by the National Natural Science Foundation of China (Grants No.~11704061 and No.~11974296) and Hong Kong RGC (Grants No.~16300522 and No.~16302321).

\end{acknowledgments}

\appendix

\section{Symmetry classifications}
\label{sec_1}

In Appendix~\ref{sec_1_1}, definitions of parity-time and parity-particle-hole symmetries are first presented and explained. Then, their constraints on the energy spectra are derived, especially for those non-Hermitian systems given in Appendix~\ref{sec_1_2}. Finally, we identify the symmetry classes of $H_1$ and $H_2$ in Appendix~\ref{sec_1_3}.
\par

\subsection{Parity-time and parity-particle-hole symmetries}
\label{sec_1_1}

A system characterized by Hamiltonian $H$ is said to have time-reversal symmetry (TRS) if 
\begin{equation}
\begin{gathered}
U_{\mathcal{T}}H^\ast U^{-1}_{\mathcal{T}} =H
\end{gathered}\label{eq_a_1}
\end{equation}
in the real space, or 
\begin{equation}
\begin{gathered}
u_{\mathcal{T}} h^\ast(-\bm{k})u^{-1}_{\mathcal{T}}=h(\bm{k})
\end{gathered}\label{eq_a_2}
\end{equation}
in the momentum space. Here, $U_{\mathcal{T}}U^\dagger_{\mathcal{T}}=I,u_{\mathcal{T}}u^\dagger_{\mathcal{T}}=I$. The system is said to have parity-time symmetry (PTS) if $H$ does not change under a combination of time-reversal and parity inversion transformations. For single particle Hamiltonians in the real space, the parity operator can be written as $U_{\mathcal{P}}\mathcal{P}$ with $U_{\mathcal{P}}U^\dagger_{\mathcal{P}}=I$ and $\mathcal{P}$ representing spatial inversion on lattice, i.e, $\bm{i}=(i_x,i_y)$ goes to $\bm{i}=(-i_x,-i_y)$. Consequently, the parity-inversion transformation changes $\bm{k}$ to $-\bm{k}$ for Bloch Hamiltonians in the momentum space. Then, we say a non-Hermitian system with PTS if
\begin{equation}
\begin{gathered}
U_{\mathcal{PT}}\mathcal{P}H^\ast (U_{\mathcal{PT}}\mathcal{P})^{-1} =H
\end{gathered}\label{eq_a_3}
\end{equation}
in the real space, or
\begin{equation}
\begin{gathered}
u_{\mathcal{PT}}h^\ast(\bm{k})u^{-1}_{\mathcal{PT}} =h(\bm{k})
\end{gathered}\label{eq_a_4}
\end{equation}
in the momentum space. By introducing the operator $\mathcal{K}_k$ that changes $\bm{k}$ to $-\bm{k}$, we can write Eqs.~\eqref{eq_a_2} and Eq.~\eqref{eq_a_4} in elegant forms as $[u_{\mathcal{T}}\mathcal{K}\mathcal{K}_k,h(\bm{k})]_{\zeta=1}=0$ and $[u_{\mathcal{PT}}\mathcal{K},h(\bm{k})]_{\zeta=1}=0$, respectively. 
\par

In addition to PTS, our non-Hermitian models $H_1$ and $H_2$ also have parity-particle-hole symmetry (PPHS), a symmetry relating to the product of particle-hole and parity-inversion transformations. We first define particle-hole symmetry (PHS) for non-Hermitian systems:
\begin{equation}
\begin{gathered}
U_{P} H^T U_{P}^{-1}=-H
\end{gathered}\label{eq_a_5}
\end{equation}
in the real space, or 
\begin{equation}
\begin{gathered}
u_{P} h^T(-\bm{k}) u_{P}^{-1}=-h(\bm{k})
\end{gathered}\label{eq_a_6}
\end{equation}
in the momentum space with $U_{P}U^\dagger_{P}=I$ and $u_{P}u^\dagger_{P}=I$. Similar to PTS, we say a non-Hermitian system has PPHS if
\begin{equation}
\begin{gathered}
(U_{\mathcal{P}P}\mathcal{P}) H^T (U_{\mathcal{P}P}\mathcal{P})^{-1}=-H
\end{gathered}\label{eq_a_7}
\end{equation}
in the real space, or 
\begin{equation}
\begin{gathered}
u_{\mathcal{P}P}h(\bm{k})^T u_{\mathcal{P}P}^{-1}=-h(\bm{k})
\end{gathered}\label{eq_a_8}
\end{equation}
in the momentum space. In Eqs.~\eqref{eq_a_7} and~\eqref{eq_a_8}, $U_{\mathcal{P}P}=U_{\mathcal{P}}U_{P}$ and $u_{\mathcal{P}P}=u_{\mathcal{P}}u_{P}$. Likewise, Eqs.~\eqref{eq_a_6} and \eqref{eq_a_8} can be rewritten as $[u_{P}\lambda\mathcal{K}_k,h(\bm{k})]_{\zeta=-1}=0$ and $[u_{\mathcal{P}P}\lambda,h(\bm{k})]_{\zeta=-1}=0$, respectively. Here, $\lambda$ is the transpose operator. We summarize the definition of TRS, PTS, PHS, and PPHS in Table~\ref{tab_1}.
\par

\begin{table}
\caption{\label{tab_1} Definitions of TRS, PTS, PHS, and PPHS in the real and momentum spaces. $H$ is the non-Hermitian single-particle Hamiltonian in the real space, and $h(\bm{k})$ is the non-Hermitian Bloch Hamiltonian. In the absence of disorders, $H$ can be block-diagonalized as $H=\sum_{\bm{k}}a^\dagger_{\bm{k}}h(\bm{k})a_{\bm{k}}$. $\mathbb{P}$ is the parity inversion operator. PTS (PPHS) means a non-Hermitian system is invariant under a combination of time-reversal (particle-hole) and parity-inversion transformations.}
\begin{ruledtabular}
\begin{tabular}{ccc}
symmetry & real space & momentum space \\
\hline
TRS & $U_{\mathcal{T}}H^\ast U^{-1}_{\mathcal{T}}=H$ & $u_{\mathcal{T}}h^{\ast}(-\bm{k})u^{-1}_{\mathcal{T}}=h(\bm{k})$ \\
PTS & $U_{\mathcal{PT}}\mathcal{P}H^\ast \mathcal{P}^{-1} U^{-1}_{\mathcal{PT}}=H$ & $u_{\mathcal{PT}}h^\ast(\bm{k})u^{-1}_{\mathcal{PT}}=h(\bm{k})$ \\
\hline\hline
PHS & $U_{P}H^T U^{-1}_{P}=-H$ & $u_{P}h^T(-\bm{k}) u^{-1}_{P}=-h(\bm{k})$ \\
PPHS &  $U_{\mathcal{P}P}\mathcal{P} H^T \mathcal{P}^{-1}U^{-1}_{\mathcal{P}P}=-H$ & $u_{\mathcal{P}P}h^T(\bm{k})u^{-1}_{\mathcal{P}P}=-h(\bm{k})$ \\
\end{tabular}
\end{ruledtabular}
\end{table}

Before the end of this section, we would like to emphasize that disordered non-Hermitian Hamiltonians are defined in the real space rather than the momentum space, and one can calculate the corresponding Bloch Hamiltonian only in the clean limit. A disordered non-Hermitian system's Hamiltonian with a symmetry guarantees that its Bloch Hamiltonian is also invariant with the corresponding symmetry operation, but it is not vice versa since disorders may break the symmetry. 
\par
\quad\par

\subsection{Constraints of complex energies due to symmetries}
\label{sec_1_2}

PTS gives some constraints on the complex eigenenergies. Let us assume a Bloch Hamiltonian $h(\bm{k})$ has PTS, i.e., Eq.~\eqref{eq_a_4}, and $|\epsilon\rangle$ is a right eigenstate of $h(\bm{k})$ with an eigenenergy $\epsilon$, i.e., $h(\bm{k})|\epsilon\rangle=\epsilon|\epsilon\rangle$. Then,
\begin{equation}
\begin{gathered}
u_{\mathcal{PT}}\mathcal{K}h(\bm{k}) |\epsilon\rangle=\epsilon^\ast u_{\mathcal{PT}}\mathcal{K}|\epsilon\rangle=h(\bm{k}) u_{\mathcal{PT}}\mathcal{K}|\epsilon\rangle.
\end{gathered}\label{eq_a_9}
\end{equation}
Equation~\eqref{eq_a_9} means that $u_{\mathcal{PT}}\mathcal{K}|\epsilon\rangle$ is also a right eigenstate of $h(\bm{k})$ with eigenenergy $\epsilon^\ast$. If the two states $|\epsilon\rangle$ and $u_{\mathcal{PT}}\mathcal{K}|\epsilon\rangle$ are the same, $\epsilon=\epsilon^\ast$, i.e., $\epsilon$ is real. This can happen for either $u_{\mathcal{PT}}u^\ast_{\mathcal{PT}}=I$ or $u_{\mathcal{PT}}u^\ast_{\mathcal{PT}}=-I$ if $h(\bm{k})$ has a double degeneracy~\cite{cwang_prbl_2022}. 
\par

PPHS also gives constraints on the complex eigenenergies. Recall that PPHS is defined as $[u_{C}\lambda,h(\bm{k})]_{\zeta=-1}=0$ in the momentum space, where $\lambda$ is the transpose operator satisfying $\lambda(cA)|\alpha\rangle=cA^T|\alpha\rangle$ with $c,A,|\alpha\rangle$ being arbitrary complex number, operator, and ket, respectively. For a corresponding left eigenstate of $|\epsilon\rangle$ satisfying $h^\dagger(\bm{k})|\tilde{\epsilon}\rangle=\epsilon^\ast|\tilde{\epsilon}\rangle$,
\begin{equation}
\begin{gathered}
u^\ast_{\mathcal{P}P} h^\dagger(\bm{k})|\tilde{\epsilon}\rangle=u^\ast_{\mathcal{P}P} \epsilon^\ast |\tilde{\epsilon}\rangle=-h^\ast(\bm{k})u^\ast_{\mathcal{P}P} |\tilde{\epsilon}\rangle.
\end{gathered}\label{eq_a_10}
\end{equation}
To derive Eq.~\eqref{eq_a_10}, we have used $u^\ast_{\mathcal{P}P} h^\dagger(\bm{k})=-h^\ast(\bm{k})u^\ast_{\mathcal{P}P} $ by taking the complex conjugate of Eq.~\eqref{eq_a_8}. Multiply $\mathcal{K}$ to Eq.~\eqref{eq_a_10}:
\begin{equation}
\begin{gathered}
h(\bm{k})\left( u_{\mathcal{P}P} \mathcal{K}|\tilde{\epsilon}\rangle \right)=-\epsilon \left( u_{\mathcal{P}P} \mathcal{K}|\tilde{\epsilon}\rangle \right).
\end{gathered}\label{eq_a_11}
\end{equation}
Therefore, there always exists a right eigenstate $u_{\mathcal{P}P} \mathcal{K}|\tilde{\epsilon}\rangle$ with energy $-\epsilon$. Namely, PPHS makes the complex-energy spectrum symmetric to the origin of the complex-energy plane. 
\par

\subsection{Symmetry classification}
\label{sec_1_3}

Altland-Zirnbauer (AZ) classification is a well-established approach to determine the symmetry class of a non-Hermitian system~\cite{kkawabata_prx_2019}. Noticeably, our models $H_1$ and $H_2$ go beyond the AZ classification due to the presence of PTS and PPHS shown in Table~\ref{tab_1}. However, we would like to determine the symmetry classes of our models within the framework of AZ classification. TRS and PHS are two symmetries in the AZ classification. In addition, one require time-reversal symmetry$^\dagger$ (TRS$^\dagger$), particle-hole symmetry$^\dagger$ (PHS$^\dagger$), chiral symmetry (CS), and sub-lattice symmetry (SLS). TRS$^\dagger$ is defined as
\begin{equation}
\begin{gathered}
\left\{
\begin{array}{cc}
U_{P'}H^T U^{-1}_{P'}=H & \text{real space} \\
\quad\\
u_{P'}h^T(-\bm{k})u^{-1}_{P'}=h(\bm{k}) & \text{momentum space} \\
\end{array}
\right..
\end{gathered}\label{eq_a_12}
\end{equation}
PHS$^\dagger$ is defined as
\begin{equation}
\begin{gathered}
\left\{
\begin{array}{cc}
U_{\mathcal{T}'}H^\ast U^{-1}_{\mathcal{T}'}=-H & \text{real space} \\
\quad\\
u_{\mathcal{T}'}h^\ast(-\bm{k})u^{-1}_{\mathcal{T}'}=-h(\bm{k}) & \text{momentum space} \\
\end{array}
\right..
\end{gathered}\label{eq_a_13}
\end{equation} 
CS is defined as
\begin{equation}
\begin{gathered}
\left\{
\begin{array}{cc}
U_{\Gamma}H^\dagger U^{-1}_{\Gamma}=-H & \text{real space} \\
\quad \\
u_{\Gamma}h^\dagger (\bm{k})u^{-1}_{\Gamma}=-h(\bm{k}) & \text{momentum space} \\
\end{array}
\right.,
\end{gathered}\label{eq_a_14}
\end{equation}
and SLS is defined as
\begin{equation}
\begin{gathered}
\left\{
\begin{array}{cc}
U_{\cal{S}}H U^{-1}_{\cal{S}}=-H & \text{real space} \\
\quad \\
u_{\cal{S}}h(\bm{k})u^{-1}_{\cal{S}}=-h(\bm{k}) & \text{momentum space} \\
\end{array}
\right..
\end{gathered}\label{eq_a_15}
\end{equation}
CS and SLS can be treated as combinations of PHS and TRS$^\dagger$ and PHS and TRS, respectively. On the one hand, if none of TRS, PHS, PHS$^\dagger$, and TRS$^\dagger$ are preserved, one requires to determine whether the non-Hermitian system has CS and SLS. On the other hand, if all TRS, PHS, PHS$^\dagger$, and TRS$^\dagger$ are preserved, CS and SLS are solely determined.
\par

Let us determine the symmetry class of $H_1$. Note that
\begin{equation}
\begin{gathered}
H^\ast_1=\sum_{\bm{i}}c^\dagger_{\bm{i}}[(w_{\bm{i}}-iu_{\bm{i}})\sigma_1-i\kappa\sigma_3]c_{\bm{i}}\\
-\sum_{\bm{i}}\left[ \dfrac{i\alpha}{2}c^\dagger_{\bm{i}}(\sigma_2 c_{\bm{i}+\hat{x}}+\sigma_1 c_{\bm{i}+\hat{y}})+H.c. \right],
\end{gathered}\label{eq_a_16}
\end{equation}
\begin{equation}
\begin{gathered}
H^T_1=\sum_{\bm{i}}c^\dagger_{\bm{i}}[(w_{\bm{i}}+iu_{\bm{i}})\sigma_1+i\kappa\sigma_3]c_{\bm{i}}\\
-\sum_{\bm{i}}\left[ \dfrac{i\alpha}{2}c^\dagger_{\bm{i}}(\sigma_2 c_{\bm{i}+\hat{x}}+\sigma_1 c_{\bm{i}+\hat{y}})+H.c. \right],
\end{gathered}\label{eq_a_17}
\end{equation}
and
\begin{equation}
\begin{gathered}
H^\dagger_1=\sum_{\bm{i}}c^\dagger_{\bm{i}}[(w_{\bm{i}}-iu_{\bm{i}})\sigma_1-i\kappa\sigma_3]c_{\bm{i}} \\
-\sum_{\bm{i}}\left[ \dfrac{i\alpha}{2}c^\dagger_{\bm{i}}(\sigma_2 c_{\bm{i}+\hat{x}}-\sigma_1 c_{\bm{i}+\hat{y}})+H.c. \right].
\end{gathered}\label{eq_a_18}
\end{equation}
From Eqs.~\eqref{eq_a_16} and \eqref{eq_a_17}, one can see the hopping terms of $H^\ast_1$ and $H^T_1$ are the same since the non-Hermiticity is introduced by the on-site terms. For $w_{\bm{i}}\neq 0,u_{\bm{i}}=0$ (see Fig.~\ref{fig1}(b)), TRS, PHS, TRS$^\dagger$, PHS$^\dagger$, and SLS are broken, but CS is preserved with $U_{\Gamma}=\sigma_3\otimes I$ and $u_{\Gamma}=\sigma_3$. Here, $I$ is the unit matrix acting on the coordinate subspace. Hence, $H_1$ belongs to class AIII. For $w_{\bm{i}}\neq 0,u_{\bm{i}}\neq 0$ (see Fig.~\ref{fig1}(a)), $H_1$ belongs to class A, where TRS, PHS, TRS$^\dagger$, PHS$^\dagger$, CS, and SLS are broken. 
\par 

$H_1$ also has PTS and PPHS. Note that
\begin{equation}
\begin{gathered}
\mathcal{P}H^\ast_1\mathcal{P}^{-1}=\sum_{-\bm{i}}c^\dagger_{-\bm{i}}[(w_{-\bm{i}}-iu_{-\bm{i}})\sigma_1-i\kappa\sigma_3]c_{-\bm{i}}\\
-\sum_{-\bm{i}}\left[ \dfrac{i\alpha}{2}c^\dagger_{-\bm{i}}(\sigma_2 c_{-\bm{i}+\hat{x}}+\sigma_1 c_{-\bm{i}+\hat{y}})+H.c. \right] \\
=\sum_{\bm{i}}c^\dagger_{\bm{i}}[(w_{\bm{i}}-iu_{\bm{i}})\sigma_1-i\kappa\sigma_3]c_{\bm{i}}\\
+\sum_{\bm{i}}\left[ \dfrac{i\alpha}{2}c^\dagger_{\bm{i}}(\sigma_2 c_{\bm{i}+\hat{x}}+\sigma_1 c_{\bm{i}+\hat{y}})+H.c. \right].
\end{gathered}\label{eq_a_19}
\end{equation}
To derive Eq.~\eqref{eq_a_19}, we have reversed the $x$ and $y$ axes. Effectively, $\mathcal{P}$ keeps the on-site terms but takes the complex-conjugate of the coefficients of the hopping terms. For $w_{\bm{i}}\neq 0,u_{\bm{i}}=0$, we can choose $U_{\mathcal{PT}}=\sigma_1\otimes I$ such that Eq.~\eqref{eq_a_5} is satisfied. One can also use an elegant form of $U_{\mathcal{PT}}\mathcal{P}=\sigma_1(-1)^{i_x+i_y}$ to write the parity-time operator. Differently, one cannot find a proper $U_{\mathcal{PT}}$ for $w_{\bm{i}}\neq 0,u_{\bm{i}}\neq 0$. Hence, PTS is preserved only if $u_{\bm{i}}=0$. On the other hand,
\begin{equation}
\begin{gathered}
\mathcal{P}H^T_1\mathcal{P}^{-1}=\sum_{\bm{i}}c^\dagger_{\bm{i}}[(w_{\bm{i}}+iu_{\bm{i}})\sigma_1+i\kappa\sigma_3]c_{\bm{i}}\\
+\sum_{\bm{i}}\left[ \dfrac{i\alpha}{2}c^\dagger_{\bm{i}}(\sigma_2 c_{\bm{i}+\hat{x}}+\sigma_1 c_{\bm{i}+\hat{y}})+H.c. \right].
\end{gathered}\label{eq_a_20}
\end{equation}
One can always choose $U_{\mathcal{P}P}=i\sigma_2\otimes I$ such that PPHS is preserved. 
\par

\begin{table*}
\caption{\label{tab_2} AZ classification has six fundamental symmetries known as TRS, PHS, TRS$^\dagger$, PHS$^\dagger$, CS, and SLS. The entries 0 indicating no symmetry and $\pm 1$ meaning that $UU^\ast=\pm I$, respectively. In addition to the six fundamental symmetries, our models have two additional symmetries, known as PTS (parity-time symmetry) and PPHS (parity-particle-hole symmetry), whose definitions are given in Table~\ref{tab_1}. The entries 0 and $\pm 1$ means without and with the two symmetries ($\pm 1$ stands for $UU^\ast=\pm I$). The last column in the table is the symmetry class of Hermitian systems, whose criticality of Anderson localization transitions is reported to be equivalent to the corresponding non-Hermitian symmetry class~\cite{xluo_prresearch_2022}.}
\begin{ruledtabular}
\begin{tabular}{ccccccccccc}
\multicolumn{11}{c}{$H_1$} \\
& TRS & PHS & TRS$^\dagger$ & PHS$^\dagger$ & CS & SLS &  class  & PTS & PPHS & Hermitian class \\
\hline
$w_{\bm{i}}\neq 0,u_{\bm{i}}=0$ & 0 & 0 & 0 & 0 & 1 & 0 & AIII  & 1 & -1 & A \\
$w_{\bm{i}}\neq 0,u_{\bm{i}}\neq 0$ & 0 & 0 & 0 & 0 & 0& 0 & A & 0 & -1 & AIII \\
\hline
\multicolumn{11}{c}{$H_2$} \\
& TRS & PHS & TRS$^\dagger$ & PHS$^\dagger$ & CS & SLS & class & PTS & PPHS & Hermitian class \\
\hline
$w_{\bm{i}}\neq 0,u_{\bm{i}}=0$ & -1 & 1 & -1 & -1 & 1 & 1 & DIII$+{\cal S}_{-+}$ & 1 & -1 & AII \\
$w_{\bm{i}}\neq 0,u_{\bm{i}}\neq 0$ & 0 & 1 & -1 & 0 & 0 & 1  & DIII$+{\cal S}_{-}$ & 0 & -1 & DIII \\
\end{tabular}
\end{ruledtabular}
\end{table*}  

Now, let us turn to $H_2$. Note that
\begin{equation}
\begin{gathered}
H^\ast_2=\sum_{\bm{i}}c^\dagger_{\bm{i}}[-(w_{\bm{i}}-i u_{\bm{i}})\tau_2\sigma_0-i\kappa\tau_3\sigma_3]c_{\bm{i}}\\
-\sum_{\bm{i}}\left[ \dfrac{i\alpha}{2}c^\dagger_{\bm{i}}(\tau_0\sigma_2 c_{\bm{i}+\hat{x}}+\tau_0\sigma_1 c_{\bm{i}+\hat{y}})+H.c. \right],
\end{gathered}\label{eq_a_21}
\end{equation}
and
\begin{equation}
\begin{gathered}
H^T_2=\sum_{\bm{i}}c^\dagger_{\bm{i}}[-(w_{\bm{i}}+i u_{\bm{i}})\tau_2\sigma_0+i\kappa\tau_3\sigma_3]c_{\bm{i}}\\
-\sum_{\bm{i}}\left[ \dfrac{i\alpha}{2}c^\dagger_{\bm{i}}(\tau_0\sigma_2 c_{\bm{i}+\hat{x}}+\tau_0\sigma_1 c_{\bm{i}+\hat{y}})+H.c. \right].
\end{gathered}\label{eq_a_22}
\end{equation}
For $w_{\bm{i}}\neq 0,u_{\bm{i}}=0$ (see Fig.~\ref{fig1}(c)), TRS, PHS, TRS$^\dagger$, and PHS$^\dagger$ are preserved with $U_{\mathcal{T}}=(i\tau_3\sigma_2)\otimes I,U_{P}=(\tau_0\sigma_1)\otimes I,U_{P'}=(i\tau_1\sigma_2)\otimes I,U_{\mathcal{T}'}=(i\tau_2\sigma_1)\otimes I$. Now, $H_2$ belongs to class DIII$+{\cal S}_{-+}$. On the other hand, for $w_{\bm{i}}\neq 0,u_{\bm{i}}\neq 0$ (see Fig.~\ref{fig1}(d)), TRS and PHS$^\dagger$ are broken since the on-site terms are complex. In this case, $H_2$ belongs to class DIII$+{\cal S}_{-}$. 
\par

$H_2$ also has PTS and PPHS. Note that
\begin{equation}
\begin{gathered}
\mathcal{P}H^\ast_2\mathcal{P}^{-1}=\sum_{\bm{i}}c^\dagger_{\bm{i}}[-(w_{\bm{i}}-i u_{\bm{i}})\tau_2\sigma_0-i\kappa\tau_3\sigma_3]c_{\bm{i}}\\
+\sum_{\bm{i}}\left[ \dfrac{i\alpha}{2}c^\dagger_{\bm{i}}(\tau_0\sigma_2 c_{\bm{i}+\hat{x}}+\tau_0\sigma_1 c_{\bm{i}+\hat{y}})+H.c. \right],
\end{gathered}\label{eq_a_23}
\end{equation}
and
\begin{equation}
\begin{gathered}
\mathcal{P}H^T_2 \mathcal{P}=\sum_{\bm{i}}c^\dagger_{\bm{i}}[-(w_{\bm{i}}+i u_{\bm{i}})\tau_2\sigma_0+i\kappa\tau_3\sigma_3]c_{\bm{i}}\\
+\sum_{\bm{i}}\left[ \dfrac{i\alpha}{2}c^\dagger_{\bm{i}}(\tau_0\sigma_2 c_{\bm{i}+\hat{x}}+\tau_0\sigma_1 c_{\bm{i}+\hat{y}})+H.c. \right].
\end{gathered}\label{eq_a_24}
\end{equation}
For $H_2$ of $w_{\bm{i}}\neq 0,u_{\bm{i}}=0$, PTS and PPHS are preserved with $U_{\mathcal{PT}}=(\tau_3\sigma_1)\otimes I$ and $U_{\mathcal{P}P}= (i\tau_0\sigma_2)\otimes I$. For $w_{\bm{i}}\neq 0,u_{\bm{i}}\neq 0$, PTS is broken but PPHS is still preserved. We summarize the presence/absence of PTS and PPHS for $H_1$ and $H_2$ in Table~\ref{tab_2}.
\par

Before the end of this section, we want to mention that $H_2$ of $u_{\bm{i}}=0$ also has pseudo-Hermitian symmetry which is defined as
\begin{equation}
\begin{gathered}
U_q H^\dagger U^{-1}_q =H
\end{gathered}\label{eq_a_25}
\end{equation}
in the real space, or
\begin{equation}
\begin{gathered}
u_q h^\dagger(\bm{k}) u^{-1}_q=h(\bm{k})
\end{gathered}\label{eq_a_26}
\end{equation}
in the momentum space. Here, $U_qU^\dagger_q=I$, and $u_q u^\dagger_q=I$. One can choose $U_q= (i\tau_2\sigma_0)\otimes I$ to satisfy Eq.~\eqref{eq_a_25}. Pseudo-Hermitian symmetry is equivalent to the AZ or AZ$^\dagger$ classes with SLS, see Ref.~\cite{kkawabata_prx_2019} for more details. Therefore, $H_1$ does not have pseudo-Hermitian symmetry due to the absence of SLS.
\par

\section{Nearest-neighbor levels to EPs}
\label{sec_2}

\begin{figure}[htbp]
\centering
\includegraphics[width=0.45\textwidth]{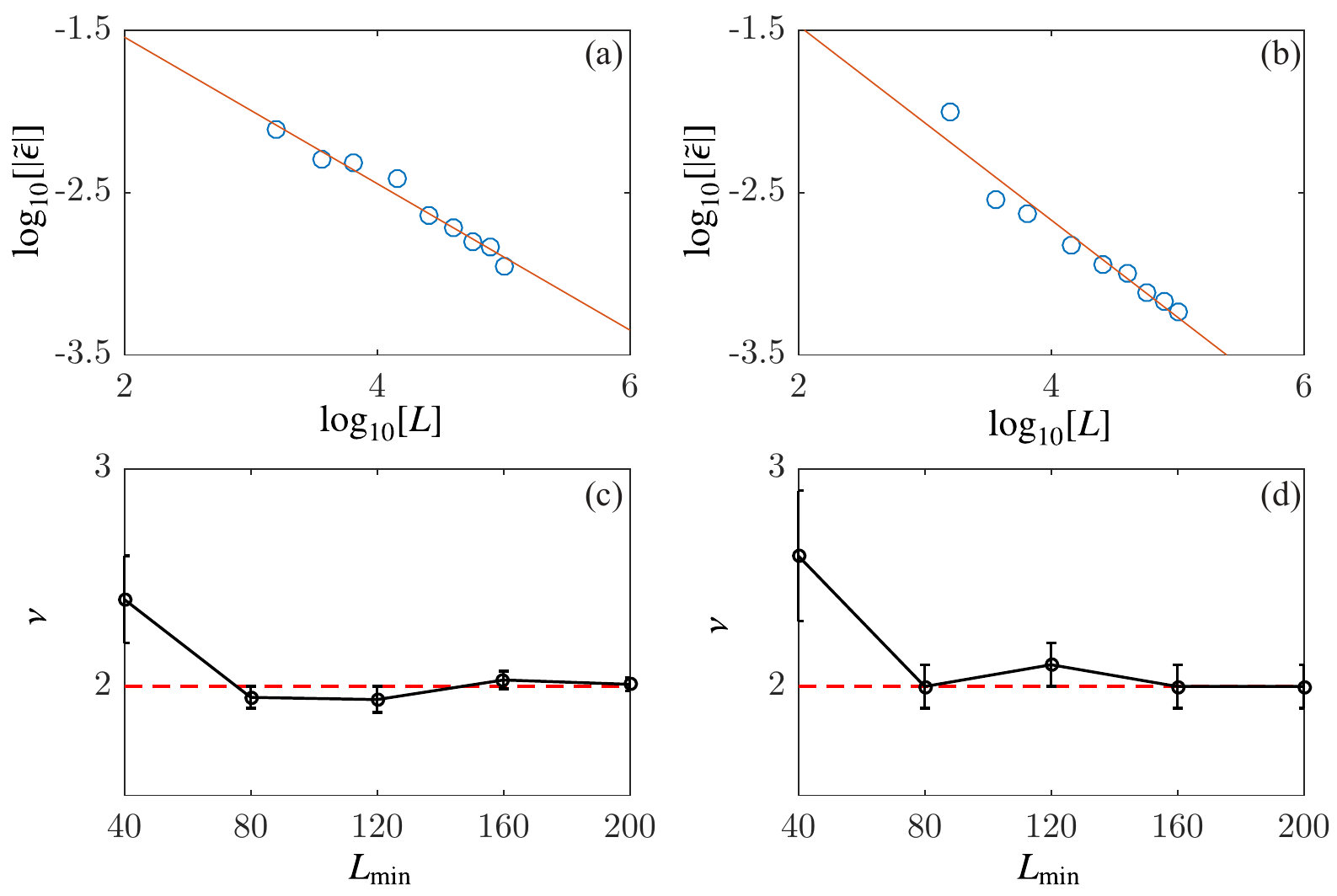}
\caption{(a) $\log_{10}[|\tilde{\epsilon}|]$ of $H_1$ as a function of $\log_{10}[L]$. The red line is a fit of $\log_{10}[|\tilde{\epsilon}|]=-\gamma \log_{10}[L]+\delta$ with $\gamma=0.45$ and $\delta=-0.64$. (b) Same as (a) but for $H_2$. Here, $\gamma=0.60$, and $\delta=-0.27$. $10^2$ samples are used here. Critical exponents for $\epsilon=0$ as a function of $ L_{\text{min}}$ obtained by $L\in [L_{\text{min}},L_{\text{max}}]$ with $L_{\text{max}}=400$. Other parameters are the same as those in Figs.~\ref{fig2}(a,b) and ~\ref{fig3}(a,b). Dashed lines in (c,d) locate $\nu=2$.}
\label{figa1}
\end{figure}

Since the non-Hermitian Hamiltonians cannot be exactly diagonalized at the EPs, we find the nearest-neighbor level $\tilde{\epsilon}$ to the EPs (locating at $\epsilon=0$ for our models $H_1$ and $H_2$) and treat the participation ratio of $\tilde{\epsilon}$, say $p_2(\tilde{\epsilon})$, as $p_2(0)$ numerically. Here, we show that $\tilde{\epsilon}$ become extremely close to the EP for $L\to\infty$ such that, for large enough sizes, $p_2(\tilde{\epsilon})\simeq p_2(0)$. To support this argument, we plot the ensemble-average $\log_{10}[|\tilde{\epsilon}|]$ as a function of system size $\log_{10}[L]$ for $H_1$ and $H_2$ with $W=0.1$ and $u_{\bm{i}}=0$, see Figs.~\ref{figa1}(a) and (b), respectively. Since the EP locates at $\epsilon=0$, $|\tilde{\epsilon}|$ is the distance between the EP and its nearest-neighbor level. As we can see, $|\tilde{\epsilon}|$ scales with $L$ as $|\tilde{\epsilon}|\sim L^{-\gamma}$ with $\gamma=0.45$ for $H_1$ and $\gamma=0.60$ for $H_2$, i.e., in both cases, $\lim_{L\to\infty}\tilde{\epsilon}= 0$.
\par

Our numerical results also show that one cannot use very small system sizes $L$ to determine the critical exponents of EPs of $H_1$ and $H_2$, i.e., a large enough size to ensure $\tilde{\epsilon}$ close to the EPs. We find that the critical exponent $\nu$ decreases with $L$ for relatively small size but becomes constant upon a critical length $\tilde{L}$. Figures~\ref{figa1}(c,d) show the critical exponents $\nu$ for $\epsilon=0$ as a function of $ L_{\text{min}}$ obtained by finite-size scaling analyses of sizes $\{ L_{\text{min}},L_{\text{min}}+40,L_{\text{min}}+80,\cdots,L_{\text{max}}\}$ with $L_{\text{max}}=400$. It is seen that the critical exponent approaches to $2$ within numerical errors for $L_{\text{min}}=\tilde{L}=80$.
\par 

\section{Finite-size scaling analysis and fitting parameters}
\label{sec_3}

To determine the critical exponent $\nu$, a finite-size scaling analysis of participation ratio $p_2$ is required. Near the critical disorder $W_c$, $p_2$ scales with the system size $L$ as
\begin{equation}
\begin{gathered}
p_2=L^D [f(L/\xi)+\phi L^y\tilde{f}(L/\xi)]
\end{gathered}\label{eq_s_2_1}
\end{equation}
with the correlation length $\xi$ diverging at $W_c$. We expand the unknown scaling function $f(x)$ and $\tilde{f}(x)$ as
\begin{equation}
\begin{gathered}
f(L/\xi)=a_0+(L/\xi)^{1/\nu}+a_1(L/\xi)^{2/\nu}, \\
\tilde{f}(L/\xi)=a_2+(L/\xi)^{1/\nu}+a_3(L/\xi)^{2/\nu}
\end{gathered}\label{eq_s_2_2}
\end{equation}
with 
\begin{equation}
\begin{gathered}
\xi=(b_1|W-W_c|+b_2|W-W_c|^2)^{-\nu}.
\end{gathered}\label{eq_s_2_3}
\end{equation}
The fitting parameters are $D,\nu,W_c,\phi,y,a_0,a_1,a_2,a_3,b_1,b_2$ (in total, there are $11$ fitting parameters) with $D$ being the fractal dimension, $\nu$ being the critical exponent of correlation length, $W_c$ being the critical disorder, and $y$ being the exponent of the irrelevant variable. The maximal likelihood estimate of the fitting parameters is obtained by minimizing the following quantity:
\begin{equation}
\begin{gathered}
\chi^2=\sum^I_{i=1}\sum^J_{j=1}\left(\dfrac{p_2(i,j)-\tilde{p}_2(i,j)}{\sigma(i,j)} \right)^2.
\end{gathered}\label{eq_s_2_4}
\end{equation}
Here, $p_2(i,j)=p_2(W_i,L_j)$ are the numerical data, and $\tilde{p}_2(i,j)$ are given by the scaling function Eq.~\eqref{eq_s_2_1}. $\chi^2$ is known as the chi-square. The total degree of freedom for a fitting process is $N_f=I\times J-M$. We also estimate the so-called goodness-of-fit $Q$ by following the standard scenario, which can be used to judge whether the fitting is acceptable or not. Generally speaking, a wrong model will often been rejected with a very small values of $Q$; while it is acceptable if $Q>10^{-3}$~\cite{wchen_prb_2019}.
\par

\begin{table*}
\caption{\label{parameter1} Chi-square fitting results for the parameters of Anderson localization transitions (ALTs) in $H_1$.}
\begin{ruledtabular}
\begin{tabular}{cccccccc}
disorders & $\epsilon$ & $W_c$ & $\nu$ & $D$ & $y$ & $N_f$ & $Q$\\
\hline
$w_{\bm{i}}\neq 0,u_{\bm{i}}=0$ & 0 & $0.85\pm 0.05$ & $1.97\pm 0.07$ & $1.0\pm 0.1$ & $-0.9\pm 0.1$ & 139 & 0.1 \\
$w_{\bm{i}}\neq 0,u_{\bm{i}}=0$ & -0.2 & $0.93\pm 0.05$ & $2.3\pm 0.1$ & $0.78\pm 0.05$ & $-1.1\pm 0.1$ & 139 & 0.2 \\
$w_{\bm{i}}\neq 0,u_{\bm{i}}=0$ & 0.08i & $0.83\pm 0.02$ & $2.32\pm 0.02$ & $0.81\pm 0.02$ & $-1.3\pm 0.1$ & 139 & 0.1 \\
\hline
$w_{\bm{i}}\neq 0,u_{\bm{i}}\neq 0$ & -0.2+0.01i & $0.62\pm 0.03$ & $2.79\pm 0.02$ & $0.67\pm 0.05$ & $-0.7\pm 0.2$ & 127 & 0.05 
\end{tabular}
\end{ruledtabular}
\end{table*}

\begin{table*}
\caption{\label{parameter2} Fitting parameters of ALTs in $H_2$.}
\begin{ruledtabular}
\begin{tabular}{cccccccc}
disorders & $\epsilon$ & $W_c$ & $\nu$ & $D$ & $y$ & $N_f$ & $Q$\\
\hline
$w_{\bm{i}}\neq 0,u_{\bm{i}}=0$ & 0 & $0.63\pm 0.08$ & $2.0\pm 0.1$ & $1.06\pm 0.03$ & $-1.5\pm 0.1$ & 121& 0.08 \\
$w_{\bm{i}}\neq 0,u_{\bm{i}}=0$ & -0.2 & $0.45\pm 0.01$ & $2.79\pm 0.05$ & $0.6\pm 0.1$ & $-1.3\pm 0.2$ & 139 & 0.03 \\
\hline 
$w_{\bm{i}}\neq 0,u_{\bm{i}}\neq 0$ & -0.2+0.01i & $0.26\pm 0.01$ & $2.56\pm 0.03$ & $0.77\pm 0.03$ & $-1.0\pm 0.1$ & 139 & 0.1 \\
\end{tabular}
\end{ruledtabular}
\end{table*}

\begin{table*}
\caption{\label{parameter3} Fitting parameters of ALTs in $H_1$ and $H_2$ with periodic boundary conditions.}
\begin{ruledtabular}
\begin{tabular}{ccccccccc}
& disorders & $\epsilon$ & $W_c$ & $\nu$ & $D$ & $y$ & $N_f$ & $Q$\\
\hline
$H_1$ & $w_{\bm{i}}\neq 0,u_{\bm{i}}=0$ & 0 & $0.81\pm 0.02$ & $2.0\pm 0.1$ & $1.0\pm 0.1$ & $-1.3\pm 0.3$ & 139 & 0.3 \\
$H_2$ & $w_{\bm{i}}\neq 0,u_{\bm{i}}=0$ & 0 & $0.37\pm 0.03$ & $2.01\pm 0.08$ & $0.97\pm 0.05$ & $-0.7\pm 0.2$ & 133 & 0.1
\end{tabular}
\end{ruledtabular}
\end{table*}

\begin{table*}
\caption{\label{parameter4} Fitting parameters of ALTs in $\tilde{H}_1$ and $\tilde{H}_2$.}
\begin{ruledtabular}
\begin{tabular}{ccccccccc}
& disorders & $\epsilon$ & $\sigma_c$ & $\nu$ & $D$ & $y$ & $N_f$ & $Q$\\
\hline
$\tilde{H}_1$ & $w_{\bm{i}}\neq 0,u_{\bm{i}}=0$ & 0 & $0.25\pm 0.03$ & $2.03\pm 0.06$ & $1.0\pm 0.1$ & $-1.0\pm 0.1$ & 139 & 0.1 \\
$\tilde{H}_2$ & $w_{\bm{i}}\neq 0,u_{\bm{i}}=0$ & 0 & $0.28\pm 0.06$ & $2.0\pm 0.1$     & $0.8\pm 0.2$ & $-1.2\pm 0.1$ & 139 & 0.02\\
\end{tabular}
\end{ruledtabular}
\end{table*}

We summarize some fitting parameters in Tables~\ref{parameter1} and \ref{parameter2}, where some of them are mentioned in Sec~\ref{section4}, together with the degree of freedom $N_f$ and the goodness-of-fit $Q$. Tables~\ref{parameter3} and \ref{parameter4} give fitting parameters of Anderson localization transitions discussed in Appendices.
\par

\section{More evidence for the universality}
\label{sec_4}

Here, we give more data to support the universality at EPs. In Appendix~\ref{sec_4_1}, we study the ALTs of $H_1$ with $u_{\bm{i}}=0$ by calculating the dimensionless conductances based on the transfer matrix method~\cite{amacKinnon_zpb_1983}. The obtained critical exponent is identical to that by participation ratios, a strong support to the universality of ALTs in Figs.~\ref{fig2}(a)-(d). In Appendix~\ref{sec_4_2}, we numerically study the ALTs in the symmetry-broken phases. In Appendices~\ref{sec_4_3} and \ref{sec_4_4}, we study the ALTs under a different boundary condition and with a different form of disorders. All ALTs near the EPs in Appendices~\ref{sec_4_2}, \ref{sec_4_3}, and \ref{sec_4_4} have the same criticality.
\par

\subsection{Dimensionless conductance}
\label{sec_4_1}

\begin{figure}[htbp]
\centering
\includegraphics[width=0.45\textwidth]{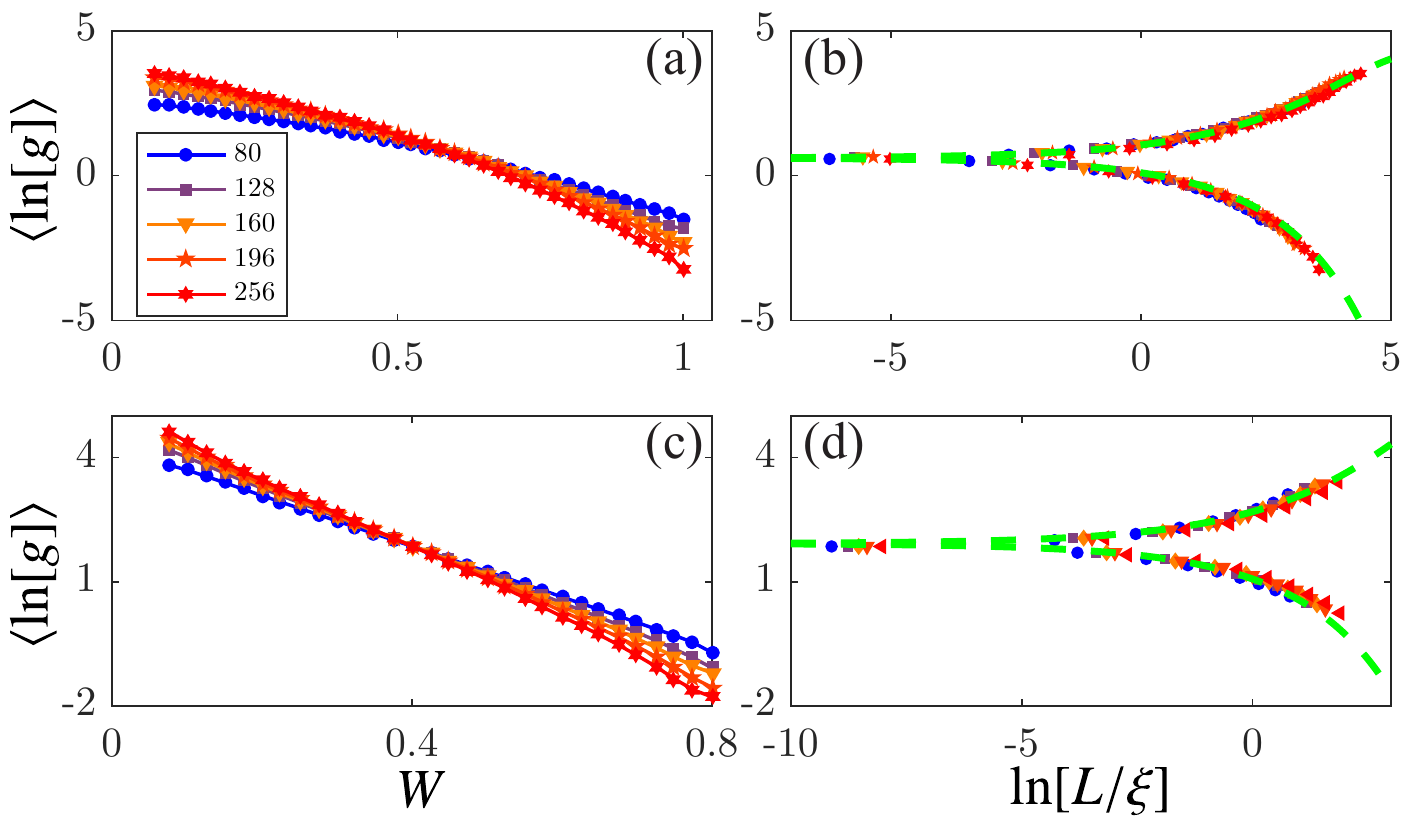}
\caption{(a) $\langle \ln [g]\rangle$ as a function of $W$ for $L=80,128,160,196,256$ at $\epsilon=-0.01$ (near the exception point). (b) Scaling function $\ln[f(x=\ln[L/\xi])]$ for (a). (c,d) Same as (a,b) but for $\epsilon=-0.2$ (in the symmetry-preserved phase).}
\label{figa2}
\end{figure}

As a self-consistent check, we investigate the localization properties of states of $H_1$ with $w_{\bm{i}}\neq0,u_{\bm{i}}=0$ through the data of dimensionless conductance. By using the transfer matrix method~\cite{amacKinnon_zpb_1983}, we calculate the dimensionless conductance of a disordered sample modelled by Eq.~\eqref{eq1} between two clean semi-infinite leads by Eq.~\eqref{eq1} with $w_{\bm{i}}=u_{\bm{i}}=0$ at a complex Fermi level $\epsilon$, i.e., $g_L=\text{Tr}[TT^\dagger]$ with $T$ being the transmission matrix. As the standard paradigm, the contact resistance is eliminated. For a given disorder $W$, the localization nature of a state at $\epsilon$ are determined by the following criteria: (i) $g_L(W)$ increases (decreases) with the size $L$ for the state being extended (localized); while is size-independent for the critical state. (ii) If there exists a quantum phase transition at a critical disorder $W_c$, $g_L(W)$ of different size $L$ near $W_c$ collapse into a smooth scaling curve $f(x=L/\xi)$ with the correlation length $\xi$ diverging as $\xi\sim |W-W_c|^{-\nu}$. 
\par

The ensemble-average $\ln [g_L]$ as a function of $W$ for $\epsilon=-0.01$ (near the exceptional point) and $-0.2$ (in the symmetry-preserved phase) for $\alpha=0.2,\kappa=0.01$ are displayed in Figs.~\ref{figa2}(a,c), respectively. It should be noted that it is unreasonable to choose $\epsilon=0$ in the transfer matrix approach since $H_1$ cannot be diagonalized at the exception point~\cite{wdheiss_jpa_2012}. Instead, we choose a state at $\epsilon=-0.01$, which is very close to the EP, and find a transition from extended states to localized states at $W_c=0.63\pm0.08$. Finite-size scaling analysis gives $\nu=2.05\pm 0.07$, which equals to $\nu=1.95\pm 0.05$ from the data of participation ratios within numerical errors. The details of the finite-size scaling analyses can be found in the Appendix of Ref.~\cite{wchen_prb_2019}. On the other hand, for $\epsilon=-0.2$ (in the symmetry-preserved phase), we also see a critical point at $W_c=0.39\pm 0.02$ near which $\xi\sim|W-W_c|^{-\nu}$ with $\nu=2.33\pm 0.05$, which equals to $\nu=2.3\pm 0.1$ shown in Fig.~\ref{fig2}(d). Hence, the obtained critical exponents $\nu$ by transfer matrix methods are consistent as those from participation ratios, which should be strong supports to the universality. 
\par

\subsection{ALT in the symmetry-broken phase}
\label{sec_4_2}

\begin{figure}[htbp]
\centering
\includegraphics[width=0.45\textwidth]{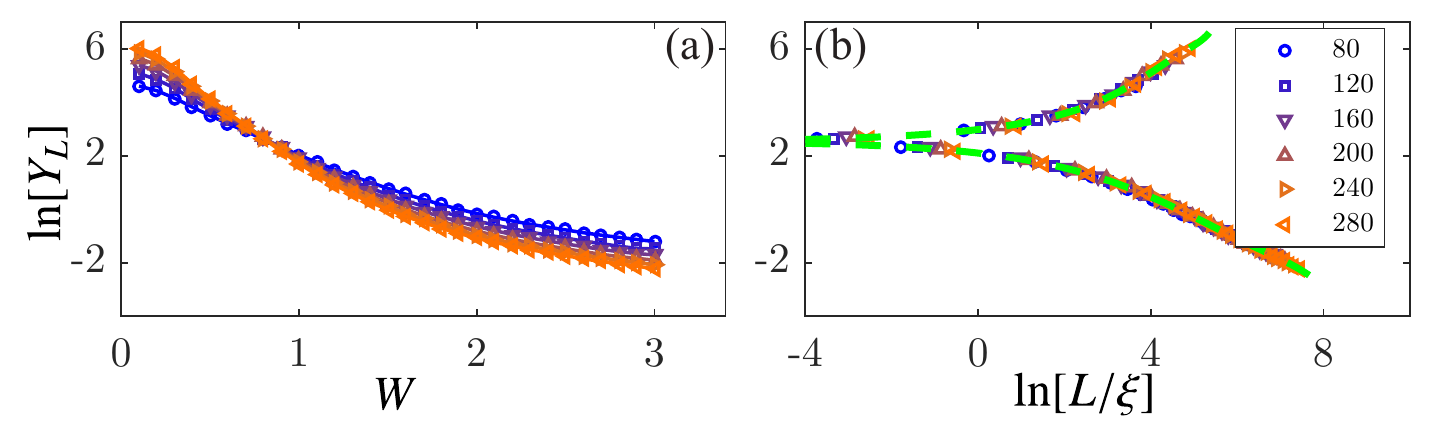}
\caption{(a) Ensemble-average $\ln [Y_L(W)]$ of $H_1$ at $\epsilon=0.08i$ (in the symmetry-broken phase) for $L=80,120,\cdots,280$. The other parameters are $\alpha=0.1,\kappa=0.1,u_{\bm{i}}=0$. $10^2$ samples are used here. (b) Scaling function $\ln [Y_L]=\ln [f(x=L/\xi)]$ for data in (a).}
\label{figa3}
\end{figure}

For a full picture for the localization nature of states of the non-Hermitian model $H_1$, we also calculate the ensemble average $\ln [Y_L]$ as a function of $W$ for various $L$, as we do in Figs.~\ref{fig2}(a)-(d), but for one state in the symmetry-broken phase ($\epsilon=0.08i$). The obtained data are displayed in Fig.~\ref{fig3}. Clearly, data for different sizes cross at one point $W_c=0.83\pm 0.02$ and those of $W<W_c$ ($W>W_c$) increases (decreases) with the system size $L$. These features indicate a transition from a band extended states to a band of localized states at $W_c$. ALTs can not only happen near the EP and in the symmetry-preserved phase as shown in Figs.~\ref{fig2}(a)-(d) but also in the symmetry-broken phase, where the eigenenergies come as the complex-conjugate pairs $(\epsilon,\epsilon^\ast)$. 
\par

To further substantiate the criticality, we perform the finite-size scaling analysis for data in Fig.~\ref{fig3}(a). Our analysis shows that the correlation lengths $\xi$ diverge as $|W-W_c|^{-\nu}$ with $\nu=2.32\pm 0.02$, and data of $\ln [Y_L]$ as a function of $\ln [L/\xi]$ collapse to two different curves (the upper and lower branches are for extended and localized states, respectively). The critical exponent equals to that of $\epsilon=-0.2$ in the symmetry-preserved phase. As we explained in the main text, ALTs of the symmetry-preserved and symmetry-broken phases belong to the same universality class. Numerical data in  Fig.~\ref{fig2}(c,d) and Fig.~\ref{figa3} confirm this assertion.  
\par

\subsection{ALTs under periodic boundary conditions}
\label{sec_4_3}

\begin{figure}[htbp]
\centering
\includegraphics[width=0.45\textwidth]{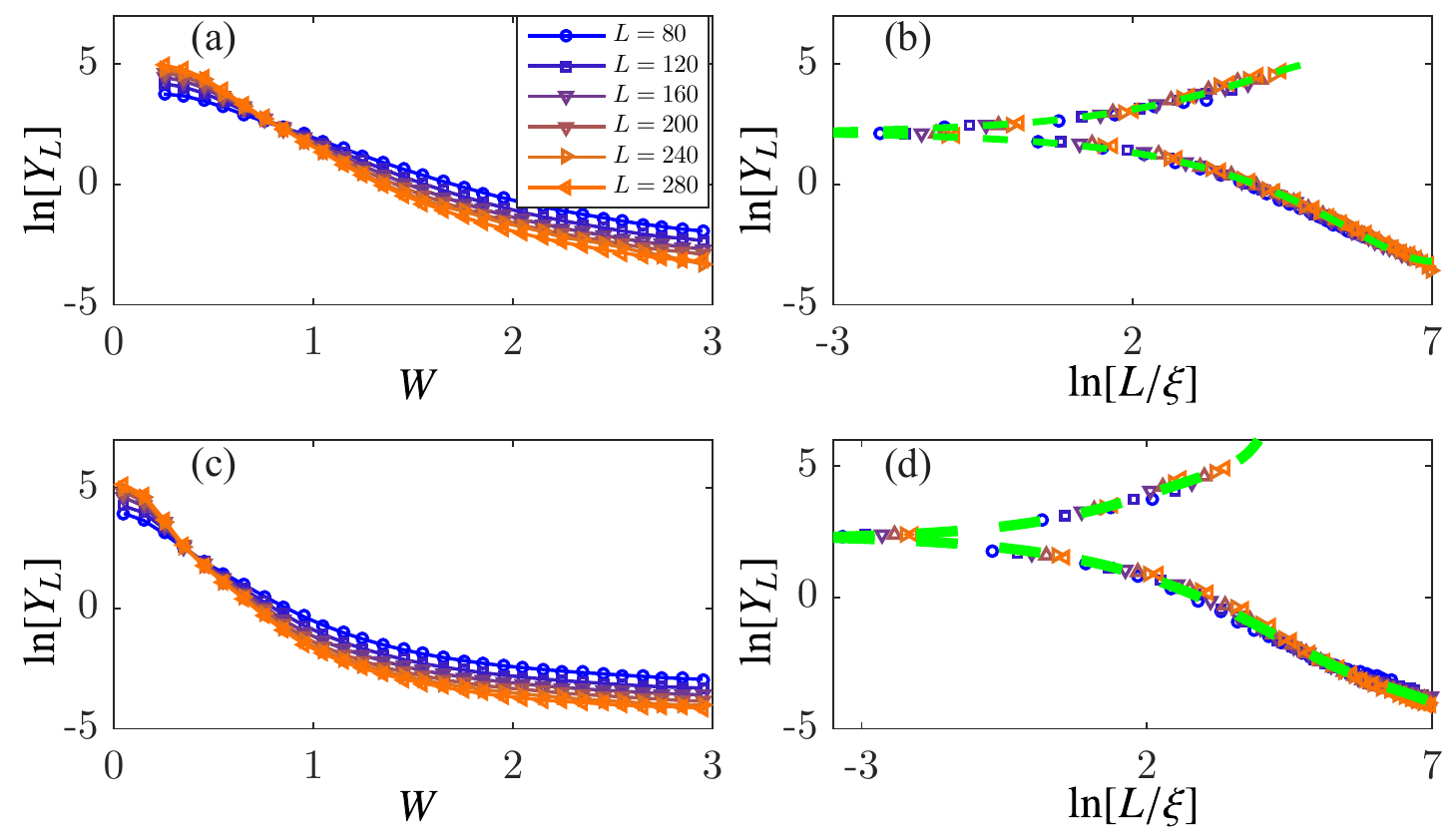}
\caption{(a) Ensemble-average $\ln[Y_L]$ as a function of $\sigma$ for $\epsilon=0$ and $L=80,120,160,200,240,280$ for $H_1$ with periodic boundary conditions. (b) Scaling function $\ln[Y_L(\ln[L/\xi])]$ of (a). (c,d) Same as (a,b) but for $H_2$. The remaining parameters are the same as those in Figs.~\ref{fig2} and \ref{fig3}. Other fitting parameters are given in Table~\ref{parameter3}.}
\label{figa4}
\end{figure}

In Figs.~\ref{fig2} and \ref{fig3}, we apply open boundary conditions to the 2D Hamiltonians $H_1$ and $H_2$. Here, we change the boundary condition to periodic boundary conditions and see whether the critical exponent $\nu$ is different for $\epsilon=0$ (EPs). The calculated $\ln [Y_L]$ as a function of $W$ for $H_1$ with $w_{\bm{i}}\neq 0,u_{\bm{i}}=0$ is depicted in Fig.~\ref{figa4}(a) with a critical disorder at $W_c=0.81\pm 0.02$. The finite size scaling analysis yields $\nu=2.0\pm 0.1$. Likewise, we find the ALT at $W_c=0.37\pm 0.03$ with the critical exponent $\nu=2.01\pm 0.08$, see Fig.~\ref{figa4}(c). Hence, the universality at EPs are not affected by choosing a different boundary condition. The scaling functions are shown in Figs.~\ref{figa4}(c) and (d). Other fitting parameters are in Table~\ref{parameter3}.
\par

\subsection{ALTs for Gaussian distributions}
\label{sec_4_4}

\begin{figure}[htbp]
\centering
\includegraphics[width=0.45\textwidth]{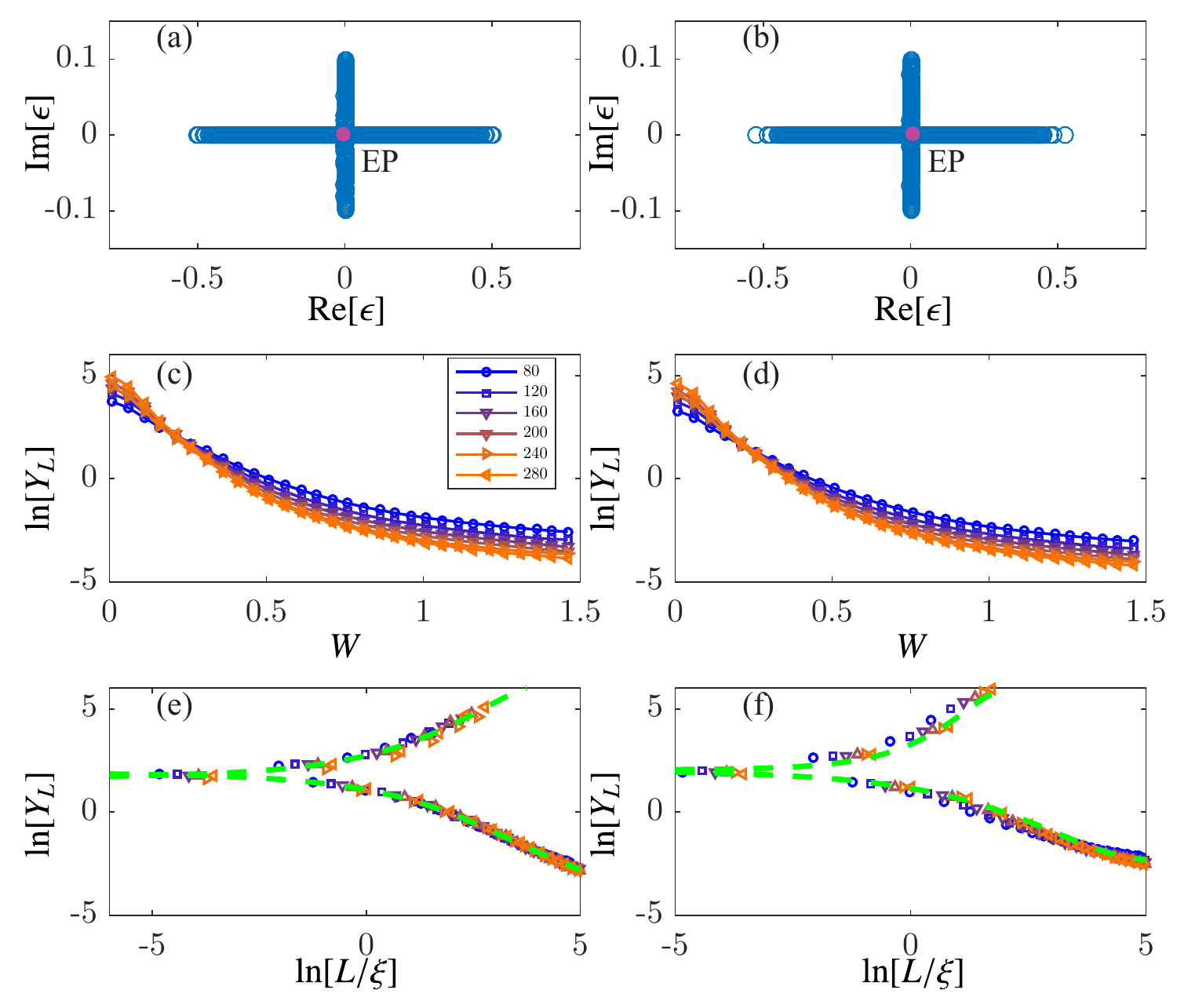}
\caption{(a,b) $\text{Re}[\epsilon]$ vs $\text{Im}[\epsilon]$ for (a) $\tilde{H}_1$ and (b) $\tilde{H}_2$. Here, $\alpha=0.2,\kappa=0.1,\sigma=0.1$, and $L=40$ (a) and 20 (b). (c,d) Ensemble-average $\ln[Y_L]$ as a function of $\sigma$ for $L=80,120,160,200,240,280$ for (a) $\tilde{H}_1$ and (b) $\tilde{H}_2$ with $\alpha=0.2,\kappa=0.1,\epsilon=0$. The remaining parameters are the same as those in Figs.~\ref{fig2} and \ref{fig3}. (e,f) Scaling function $\ln[Y_L(\ln[L/\xi])]$ of (c,d). Other fitting parameters are given in Table~\ref{parameter4}.}
\label{figa5}
\end{figure}

In Figs.~\ref{fig2} and \ref{fig3}, we model disorders of $H_1$ and $H_2$ by random on-site potentials, whose amplitudes $w_{\bm{i}}$ and $u_{\bm{i}}$ distribute independently and uniformly in a range of $[-W/2,W/2]$. Here, we supply numerical evidence to the universality at EPs for a different form of randomness, i.e., $w_{\bm{i}}$ and $u_{\bm{i}}$ are white-noise and follow the Gaussian distribution of zero mean and variance $\sigma^2$. We label the corresponding Hamiltonians as $\tilde{H}_1$ and $\tilde{H}_2$, whose Bloch Hamiltonians are the same as those of $H_1$ and $H_2$ and the disordered on-site potentials are Gaussian. Since we focus on delocalization-localization transitions at EPs, we set $u_{\bm{i}}=0$ in what follows.
\par

Similar to $H_1$ and $H_2$, eigenenergies of $\tilde{H}_1$ and $\tilde{H}_2$ form crosses in the complex-energy plane with the second-order and fourth-order EPs locating at $\epsilon=0$. These features are visualized in Figs.~\ref{figa5}(a) and (b) for $\alpha=0.2,\kappa=0.1,\sigma=0.1$. We then investigate the Anderson localization transitions at the EPs. The calculated $\ln[Y_L(\sigma)]$ are shown in Figs.~\ref{figa5}(c) and (d) for $\tilde{H}_1$ and $\tilde{H}_2$, respectively. Finite-size scaling analyse yield $\nu=2.03\pm 0.06$ for $\tilde{H}_1$ and $\nu=2.0\pm 0.1$ for $\tilde{H}_2$, which are identical to those of $H_1$ and $H_2$ within numerical errors. The scaling functions are given in Figs.~\ref{figa5}(e) and (f). 
\par

\section{Skin effect}
\label{sec_5}

\begin{figure}[htbp]
\centering
\includegraphics[width=0.45\textwidth]{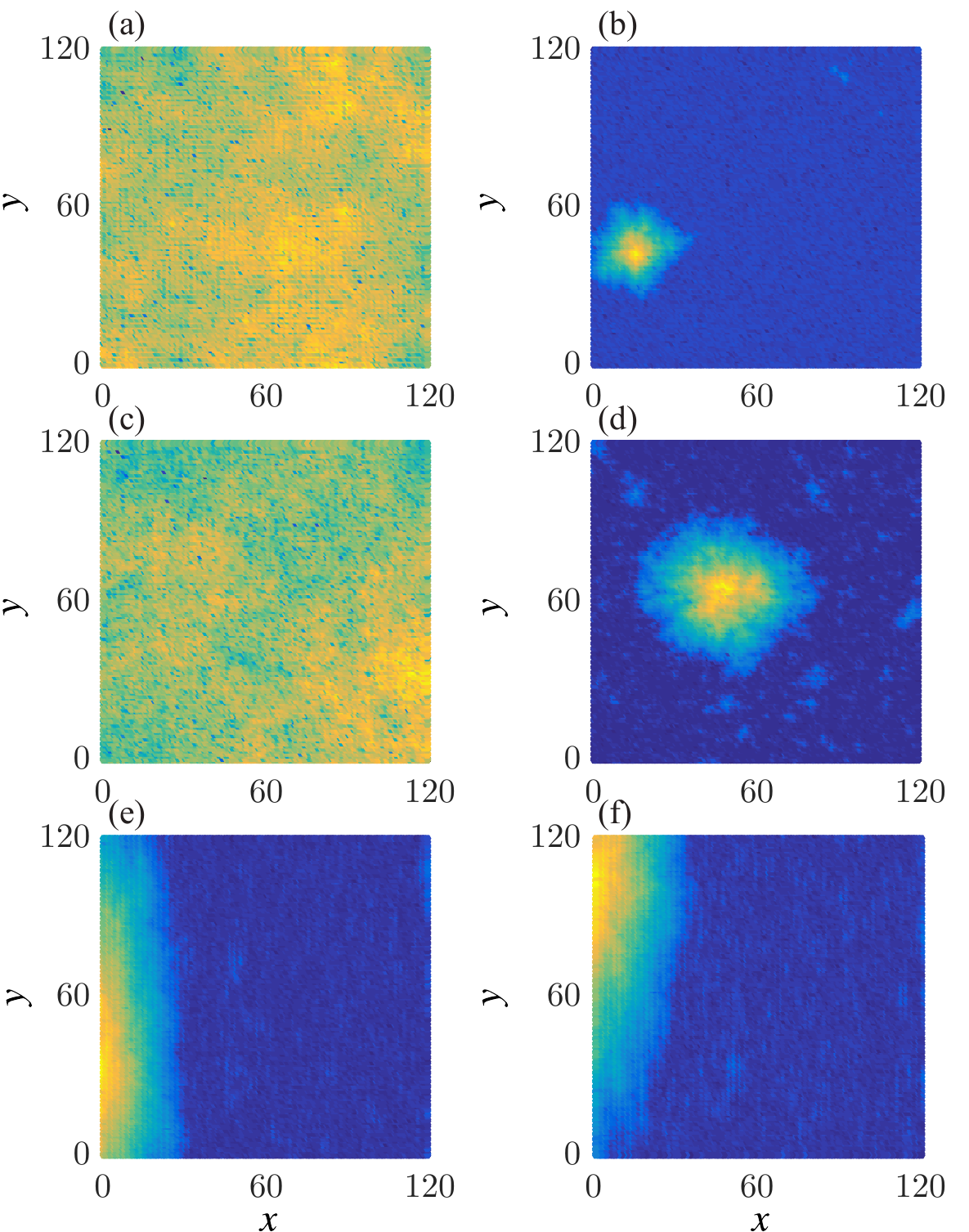}
\caption{(a) Spatial distributions of $\log_{10}|\psi_{\bm{i}}|$ of normalized wave functions of the $\epsilon=0$ state (EP) in a typical realization of $H_1$ of $w_{\bm{i}}=0,u_{\bm{i}}\neq 0$ with $\alpha=0.2,\kappa=0.1,W=0.1,L=120$. (b) Same as (a) but for $W=2.0$. (c) Same as (a) but for $H_1$ with $\alpha=0.2,\kappa=0.1,W=0.1$. (d) Same as (c) but for $W=2.0$. (e) Same as (a) but for $H_{1,\text{skin}}$ given in Eq.~\eqref{eq_s_5_4}. (f) Same as (a) but for $H_{2,\text{skin}}$ given in Eq.~\eqref{eq_s_5_5}. Colors map $\log_{10}|\psi_{\bm{i}}|$. The yellow (blue) color stands for a larger (smaller) spatial distribution.}
\label{figa6}
\end{figure}

Non-Hermiticities sometimes cause a skin effect where the wave functions localize exponentially at boundaries of systems of the open boundary conditions. This can be seen by the following low-energy continuous Hamiltonian
\begin{equation}
\begin{gathered}
h(\bm{p})=\alpha p_1\sigma_1+\alpha p_2\sigma_2+V_{\text{nh}} \\
=\alpha p_1\sigma_1+\alpha p_2\sigma_2+i\sum_{\mu=0,1,2,3,}\kappa_{\mu} \sigma_{\mu}
\end{gathered}\label{eq_s_5_1}
\end{equation}
with $\alpha$ and $\kappa_{0,1,2,3}$ being real numbers. The Hermitian part of Eq.~\eqref{eq_s_5_1} is the effective $\bm{k}\cdot\bm{p}$ Hamiltonian of the Hermitian part of model~(1) of the main text, and $V_{\text{nh}}$ is a general non-Hermitian potential. One can generalize the Hamiltonian in the real $\bm{p}$ space to that in the complex $\tilde{\bm{p}}$ space, i.e., 
\begin{equation}
\begin{gathered}
h(\tilde{p})=i\kappa_0\sigma_0+\alpha\tilde{p}_1\sigma_1+\alpha\tilde{p}_2\sigma_2+i\kappa_3\sigma_3
\end{gathered}\label{eq_s_5_2}
\end{equation}
with $\tilde{p}_{1,2}=p_{1,2}+i\kappa_{1,2}/\alpha$ being complex numbers. The role of $\kappa_{1,2}\neq 0$ can be thus seen by replacing the real wave vectors $p_{1,2}$ by the complex ones $\tilde{p}_{1,2}$ in the Bloch phase of $\exp[i\bm{p}\cdot\bm{x}]$. Hence, eigenstates of Eq.~\eqref{eq_s_5_2} localized exponentially at the boundaries if $\kappa_{1,2}\neq 0$. However, this never happens for the non-Hermitian systems with either PTS or pseudo-Hermitian symmetry. For PTS, one requires $u_{\mathcal{PT}} h^\ast(\bm{p}) u^{-1}_{\mathcal{PT}}=h(\bm{p})$. Since
\begin{equation}
\begin{gathered}
h^\ast(\bm{p})=-i\kappa_0\sigma_0+\alpha(p_1-i\kappa_1/\alpha)\sigma_1+\alpha(-p_2+i\kappa_2/\alpha)\sigma_2 \\
-i\kappa_3\sigma_3,
\end{gathered}\label{eq_s_5_3}
\end{equation}
we must set $\kappa_0=\kappa_1=\kappa_2=0$ and choose $u_{\mathcal{PT}}=\sigma_1$. Likewise, one can find that the skin effect is prohibited for a pseudo-Hermitian system.
\par

This can be seen in the following. In Fig.~\ref{figa6}(a), we show the wave function distribution $\log_{10}|\psi_{\bm{i}}(\epsilon=0)|^2$ for $H_1$, whose EPs are second-order, at a particular disorder $W=0.1$. One can see that the wave function spreads over the whole sample since it is a delocalized state. On the other hand, for a stronger disorder $W=2.0$ that is larger than $W_c=0.85$ (see Table~\ref{parameter1}), the state is highly localized at the bulk, see Fig.~\ref{figa6}(b). Similar features are also observed for $H_2$, see Figs.~\ref{figa6}(c) and (d). Noticeably, such localizations are different from that due to the skin effect. To see it, we consider the following two models:
\begin{equation}
\begin{gathered}
H_{1,\text{skin}}=\sum_{\bm{i}}c^\dagger_{\bm{i}}[(w_{\bm{i}}+iu_{\bm{i}})\sigma_1+i\kappa\sigma_2]c_{\bm{i}} \\
-\sum_{\bm{i}}\left[ \dfrac{i\alpha}{2}c^\dagger_{\bm{i}}(\sigma_2 c_{\bm{i}+\hat{x}}-\sigma_1 c_{\bm{i}+\hat{y}})+H.c. \right].
\end{gathered}\label{eq_s_5_4}
\end{equation}
and
\begin{equation}
\begin{gathered}
H_{2,\text{skin}}=\sum_{\bm{i}}c^\dagger_{\bm{i}}[(w_{\bm{i}}+i u_{\bm{i}})\tau_2\sigma_0+i\kappa\tau_0\sigma_1]c_{\bm{i}}\\ 
-\sum_{\bm{i}}\left[ \dfrac{i\alpha}{2}c^\dagger_{\bm{i}}(\tau_0\sigma_2 c_{\bm{i}+\hat{x}}-\tau_0\sigma_1 c_{\bm{i}+\hat{y}})+H.c. \right].
\end{gathered}\label{eq_s_5_5}
\end{equation}
The differences between Eqs.~\eqref{eq_s_5_4} and \eqref{eq_s_5_5} and $H_1$ and $H_2$ are the non-Hermitian on-site potentials: They are $i\kappa\sigma_2$ and $i\kappa\tau_0\sigma_1$ in Eqs.~\eqref{eq_s_5_4} and \eqref{eq_s_5_5} but $i\kappa\sigma_3$ and $i\kappa\tau_3\sigma_3$ in $H_1$ and $H_2$. The modifications of the non-Hermitian potentials lead to the skin effect in the $y$ direction such that wave functions of the two new models localized exponentially, see Figs.~\ref{figa6}(e) and (f). Such localizations are intrinsically different from Anderson localizations where wave functions can be localized at the bulk, rather than the edges, of a non-Hermitian system.
\par

\section{Honeycomb laser cavity network}
\label{sec_6}

One possibility is to model Haldane Hamiltonian by the laser cavity network as recently done in experiments that realized chiral edge states~\cite{gharari_science_2018,mabandres_science_2018}. In an early work~\cite{cwang_prb_2022}, we derived the effective Hamiltonian of coupled laser cavities on a honeycomb lattice
\begin{equation}
\begin{gathered}
H=\sum_{\bm{i} \in A}\epsilon_{a,\bm{i}} a^\ast_{\bm{i}} a_{\bm{i}} +\sum_{\bm{i} \in B}\epsilon_{b,\bm{i}} b^\ast_{\bm{i}} b_{\bm{i}}\\
+\sum_{\langle \bm{i,j}\rangle} (t_1 a^\ast_{\bm{i}} b_{\bm{j}}+c.c)+\sum_{\langle\langle \bm{i,j} \rangle\rangle}t_2e^{i\phi}(a^\ast_{\bm{i}} a_{\bm{j}}+b^\ast_{\bm{i}} b_{\bm{j}}+c.c)
\end{gathered}\label{eq_a6_1}
\end{equation}
Here, $a_{\bm{i}}$ and $b_{\bm{i}}$ are the laser field amplitudes at site $\bm{i}$ of A and B sub-lattices, respectively. $\langle \bm{i,j}\rangle$ and $\langle\langle \bm{i,j} \rangle\rangle$ denote the nearest-neighbor sites and the next-nearest-neighbor sites, respectively. Each resonator is coupled to its nearest-neighbors sites with a real coupling constant $t_1$ and to its next-nearest-neighbor sites with a complex coupling constant $t_2e^{i\phi}$ with $\phi$ being the tunable Haldane flux parameters. The complex parameters $\epsilon_{a,\bm{i}}$ and $\epsilon_{b,\bm{i}}$ are
\begin{equation}
\begin{gathered}
\epsilon_{a,\bm{i}}=\omega_{a,\bm{i}}+i\left( \tilde{g}_a-\gamma \right)
\end{gathered}\label{eq_a6_2}
\end{equation}
and
\begin{equation}
\begin{gathered}
\epsilon_{b,\bm{i}}=\omega_{b,\bm{i}}+i\left( \tilde{g}_b-\gamma \right).
\end{gathered}\label{eq_a6_3}
\end{equation}
Real numbers $\omega_{a,\bm{i}}$ and $\omega_{b,\bm{i}}$ represent the resonance frequencies of the resonator at site $\bm{i}$ of A and B sub-lattices. The resonance frequency depend on the size and the shape of the resonator. Therefore, disorders can be introduced by setting $\omega_{a,\bm{i}}$ and $\omega_{b,\bm{i}}$ randomly. The real positive number $\gamma$ is the linear loss of a resonator. Real positive number $\tilde{g}_a$ and $\tilde{g}_b$ in Eqs.~\eqref{eq_a6_2} and \eqref{eq_a6_3} are the optical gains via stimulated emission that is inherently saturated. Hereafter, we choose 
\begin{equation}
\begin{gathered}
\tilde{g}_a=\gamma+\kappa, \tilde{g}_b=\gamma-\kappa
\end{gathered}\label{eq_a6_4}
\end{equation}
such that $\epsilon_{a,\bm{i}}=\omega_{a,\bm{i}}+i\kappa,\epsilon_{b,\bm{i}}=\omega_{b,\bm{i}}-i\kappa$.
\par 

In the absence of disorders, say $\omega_{a,\bm{i}}=\omega_{b,\bm{i}}=\omega_0$, Eq.~\eqref{eq_a6_1} can be block diagonalized in the momentum space
\begin{equation}
\begin{gathered}
H=\sum_{\bm{k}}
\begin{bmatrix}
a^\ast_{\bm{k}} & b^\ast_{\bm{k}}
\end{bmatrix}h(\bm{k})
\begin{bmatrix}
a_{\bm{k}} \\
b_{\bm{k}}
\end{bmatrix},
\end{gathered}\label{eq_a6_5}
\end{equation}
where
\begin{equation}
\begin{gathered}
h(\bm{k})=h_0\sigma_0+\bm{h}\cdot\bm{\sigma}
\end{gathered}\label{eq_a6_6}
\end{equation}
with
\begin{equation}
\begin{gathered}
\left\{
\begin{array}{ccc}
h_0 & = & 2t_2\cos\phi\sum_{i=1,2,3} \cos[\bm{k}\cdot\bm{v}_i] \\
\quad\\
h_1 & = & t_1\sum_{i=1,2,3}\cos [\bm{k}\cdot\bm{u}_i] \\
\quad\\
h_2 & = & -t_1\sum_{i=1,2,3}\sin[\bm{k}\cdot\bm{u}_i] \\
\quad\\
h_3 & = & 2t_2\sin\phi\sum_{i=1,2,3}\sin[\bm{k}\cdot\bm{v}_i]+i\kappa
\end{array}
\right..
\end{gathered}\label{eq_a6_7}
\end{equation}
Note that $\sigma_{0,1,2,3}$ stand for the unit matrix and the Pauli matrices acting on the A-B sub-lattice space. In Eq.~\eqref{eq_a6_7}, $\bm{u}_1=(\sqrt{3}/2,1/2),\bm{u}_2=(-\sqrt{3}/2,1/2),\bm{u}_3=-(\bm{u}_1+\bm{u}_2)$, and $\bm{v}_1=\bm{u}_2-\bm{u}_3,\bm{v}_2=\bm{u}_3-\bm{u}_1,\bm{v}_3=\bm{u}_1-\bm{u}_2$. The Hermitian part of Eq.~\eqref{eq_a6_7} supports chiral edge states for $t_2\neq 0$ and $-\pi<\phi<0$ and $0<\phi<\pi$. Near the two distinct corners $\bm{K}=(4\pi/(3\sqrt{3}a),0)$ and $\bm{K}'=-\bm{K}$, we can expand the Bloch Hamiltonian $h(\bm{k})$ with small $\bm{p}=(p_1,p_2)$:
\begin{equation}
\begin{gathered}
h_{\pm}(\bm{p})=-3t_2\cos[\phi]\sigma_0-\dfrac{3t_1 a}{2}(\pm p_1\sigma_1-p_2\sigma_2)\\
-(\pm 3\sqrt{3}t_2\sin[\phi]-i\kappa)\sigma_3.
\end{gathered}\label{eq_a6_8}
\end{equation}
with the subscripts $\pm$ standing for the $\bm{K}$ and $\bm{K}'$, respectively. By artificially tuning $\phi=0$, Eq.~\eqref{eq_a6_8} is invariant under parity-time operation, i.e., $[\sigma_1\mathcal{K},\tilde{h}(\bm{p})]_{\zeta=1}=0$, and supports an EP at $9t^2_1a^2(p^2_1+p^2_2)/4=\kappa^2$. 
\par

An on-site randomness like $w_{\bm{i}}\sigma_1$ can be achieved by artificially changing the resonant frequencies of the cavities. In principle, one can measure the laser field amplitudes of Eq.~\eqref{eq_a6_1}, see Ref.~\cite{mabandres_science_2018}, from which the participation ratio can be calculated $\tilde{p}_2=\sum_{\bm{i}}(|a_{\bm{i}}|^4+|b_{\bm{i}}|^4)$. The ALTs can be seen by studying the size-dependence of $\tilde{p}_2$ near the EP, and the critical exponent can be calculated through the finite-size scaling analysis. We thus expect the designed laser cavity networks are ideal platform to study the Anderson localization transitions at EPs and all conclusion given in this work can be experimentally tested based on the laser cavity networks.  
\par

\end{document}